\documentclass[11pt]{article}
\usepackage{acl}

\usepackage{times}
\usepackage{latexsym}
\usepackage[T1]{fontenc}

\usepackage[utf8]{inputenc}

\usepackage{amsmath}     
\usepackage{amssymb}     
\usepackage{algorithm}
\usepackage{algpseudocode}  
\usepackage{adjustbox}
\usepackage{amsfonts}    
\usepackage{booktabs}
\usepackage{colortbl}
\usepackage{enumitem}    
\usepackage{float}
\usepackage{csquotes}
\usepackage{graphicx}
\usepackage{hyperref}
\usepackage{inconsolata}
\usepackage{microtype}
\usepackage{multirow}
\usepackage{pgfplots}
\usepackage{pgfplotstable}
\usepgfplotslibrary{groupplots}
\pgfplotsset{compat=1.18}
\usepackage{subcaption}
\usepackage{tikz}
\usepackage{tcolorbox}
\usepackage{tabularx}
\usepackage[table]{xcolor}
\usetikzlibrary{shapes.geometric, backgrounds, calc}
\usetikzlibrary{decorations.pathmorphing}
\tcbuselibrary{skins}
\usepackage{verbatim}    

\usepackage{multirow}   
\usepackage{caption}    
\usepackage{geometry}   
\usepackage{makecell}   
\usepackage{xcolor} 

\newtcolorbox{promptbox}[1]{
    colback=gray!10,           
    colframe=black,            
    colbacktitle=black,        
    coltitle=white,            
    fonttitle=\small\sffamily\bfseries, 
    enhanced,                  
    attach boxed title to top left={xshift=0mm, yshift=0mm}, 
    boxed title style={
        sharp corners,         
        rounded corners=north, 
        frame hidden,          
    },
    title=#1,                  
    arc=5pt,                   
    outer arc=5pt,
    fontupper=\small,
    left=10pt,                 
    right=10pt,                
    top=10pt,                  
    bottom=10pt,               
    width=\textwidth,          
    boxrule=1.5pt              
}

\newtcolorbox{halfpromptbox}[1]{
    colback=gray!10,           
    colframe=black,            
    colbacktitle=black,        
    coltitle=white,            
    fonttitle=\small\sffamily\bfseries, 
    enhanced,                  
    attach boxed title to top left={xshift=0mm, yshift=0mm}, 
    boxed title style={
        sharp corners,         
        rounded corners=north, 
        frame hidden,          
    },
    title=#1,                  
    arc=5pt,                   
    outer arc=5pt,
     fontupper=\small,
    left=10pt,                 
    right=10pt,                
    top=10pt,                  
    bottom=10pt,               
    width=\columnwidth,          
    boxrule=1.5pt              
}

\setlist{topsep=2pt, itemsep=0pt, parsep=0pt, partopsep=2pt}

\newcolumntype{L}{>{\raggedright\arraybackslash}p{2.5cm}} 
\newcolumntype{P}{>{\raggedright\arraybackslash}p{12cm}} 

\tcbset{
    promptstyle/.style={
        enhanced, 
        boxrule=0.8pt, 
        colback=white,
        colframe=black,
        fonttitle=\small\bfseries, 
        colbacktitle=black,
        coltitle=white,
        sharp corners,
        left=4mm, right=4mm, top=3mm, bottom=3mm,
        toptitle=2mm, bottomtitle=2mm,
    }
}

\def\signed #1{{\leavevmode\unskip\nobreak\hfil\penalty50\hskip2em
  \hbox{}\nobreak\hfil(#1)%
  \parfillskip=0pt \finalhyphendemerits=0 \endgraf}}

\newsavebox\mybox

\newcommand{\approach}{\texttt{IRAP}}

\title{Conjecture and Inquiry: Quantifying Software Performance Requirements via Interactive Retrieval-Augmented Preference Elicitation}

\author{\bf Shihai Wang$^1$\thanks{Shihai Wang is also supervised in the IDEAS Lab.}, Tao Chen$^2$\thanks{Tao Chen is the corresponding author.} \\
$^1$ School of Computer Science and Engineering, UESTC, Chengdu, China \\
$^2$ IDEAS Lab, University of Birmingham, Birmingham, UK \\
wsh2130076635@gmail.com, t.chen@bham.ac.uk}

\begin{document}
\maketitle
\begin{abstract}
Since software performance requirements are documented in natural language, quantifying them into mathematical forms is essential for software engineering. Yet, the vagueness in performance requirements and uncertainty of human cognition have caused highly uncertain ambiguity in the interpretations, rendering their automated quantification an unaddressed and challenging problem. In this paper, we formalize the problem and propose \approach, an approach that quantifies performance requirements into mathematical functions via interactive retrieval-augmented preference elicitation. \approach~differs from the others in that it explicitly derives from problem-specific knowledge to retrieve and reason the preferences, which also guides the progressive interaction with stakeholders, while reducing the cognitive overhead. Experiment results against 10 state-of-the-art methods on four real-world datasets demonstrate the superiority of \approach~on all cases with up to $40\times$ improvements under as few as five rounds of interactions.

\end{abstract}

\section{Introduction}
Software project failures often stem from unmet behavioral requirements of the software performance~\cite{DBLP:conf/re/EckhardtVFM16,DBLP:journals/tse/SayaghKAP20}.
One example is the U.S. Health Care failure in 2013~\cite{HHS2016Healthcare}, in which the vague performance preferences from the stakeholders have not been properly quantified into precise metrics, causing the software to be unable to cope with the required load surges, leading to severe financial loss. 

While quantifying performance requirements is crucial for modeling~\cite{DBLP:conf/icse/XiangChen26,DBLP:journals/tse/GongCB25}, tuning~\cite{ye2026revealing,DBLP:conf/icse/ChenChen26,chen2024mmo}, testing~\cite{DBLP:conf/icse/MaCL25,du2025causally}, and self-adaptation~\cite{DBLP:journals/corr/abs-2501-00840,DBLP:journals/tosem/ChenLBY18}, it is a challenging task since those requirements are often documented in natural language statements, which are vague and imprecise~\cite{DBLP:conf/re/EckhardtVFM16,DBLP:journals/tosem/ChenL23a, DBLP:journals/corr/abs-2509-24694}. For example, the real-world requirement $4.3.1$ for the Puget Sound Enhancements System in the \textsc{PURE} dataset~\cite{DBLP:conf/re/FerrariSG17} is:
\begin{displayquote}
``\textit{The system should support at least 1000 concurrent users.}''
\end{displayquote}
Although there is a threshold of $1,000$ users, it remains unclear to what extent the throughput is tolerable if it drops below $1,000$, nor whether above $1,000$ is equally acceptable. Even worse, commonly the stakeholders who give the natural performance requirements cannot precisely quantify their needs without assistance due to the high uncertainty of human cognition~\cite{DBLP:conf/re/Glinz07,DBLP:books/sp/ChungNYM00}. Such an \textbf{uncertain ambiguity} is what makes quantifying the preferences in performance requirements difficult.

Manually quantifying performance requirements with domain-specific language is tedious, expensive, and prone to inconsistency, especially for large projects~\cite{DBLP:conf/icse/EckhardtVF16,DBLP:journals/re/WhittleSBCB10}. Specific and automated rule-based methods like \texttt{LQPR}~\cite{DBLP:journals/corr/abs-2511-03421} can assist in the quantification, but still, they assume that the stakeholder would give all necessary information in the requirements and hence struggles to fully handle the uncertain ambiguity therein.

Existing generic methods for preference alignment have been leveraging Large Language Models (LLMs), pairing with paradigms such as Reinforcement Learning~\cite{DBLP:conf/nips/Ouyang0JAWMZASR22} or Retrieval-Augmented Generation (RAG)~\cite{DBLP:conf/icml/BorgeaudMHCRM0L22}, some of which also contain limited human feedback. However, their key limitation is they either only work on general preferences which are too costly when used for subjective, case dependent preferences from individual stakeholder in our problem~\cite{DBLP:journals/corr/abs-2306-07402}; or they solely rely on the LLM's capability to analyze the preferences without explicitly reasoning on the unique properties in the problem, i.e., in our case these are the patterns of the performance requirements. Further, their inability to continually interact with humans for progressive feedback makes coping with the uncertain ambiguity unrealistic.





To overcome the above gaps, in this paper, we propose \approach, an approach that quantifies software performance requirements as mathematical functions via interactive retrieval-augmented preference elicitation involving the stakeholder---a typical \textit{conjecture and inquiry} loop---which is important to resolve the uncertain ambiguity therein, since even the stakeholder is uncertain about his/her true preferences at the beginning. What makes \approach~unique is that while achieving \textbf{retrieval reasoning-guided interaction} based on explicit knowledge of the problem, it also strikes on several aspects to \textbf{minimize the cognitive overhead} of the stakeholder, e.g., via suggesting a quantification that is likely similar to one's true preferences and providing intuitive questions during interaction. Specifically, our contributions are:

\begin{itemize}

    \item \textbf{Theory:} We formulate the preference reasoning problem for quantifying performance requirements based on the defined notions of \textbf{precision} and \textbf{difficulty} therein.

    \item \textbf{Quantification:} We propose retrieval-generative quantification to automatically quantify an initial draft of a given natural performance requirement via dual classification and generation based on patterns and LLM.

    \item \textbf{Reasoning:} To reduce the cognitive overhead, we propose retrieval-analogical preference reasoning that uses past examples as analogies to convert the quantification of the initial draft, hence it is closer to what is most likely to be preferred by a stakeholder.

\item \textbf{Interaction:} Using the converted/reasoned quantification as a starting point, we present a tree-based question-answering interaction with the stakeholder to tune the preferences.



\end{itemize}

To evaluate \approach, we compare it against 10 state-of-the-art methods under four real-world datasets of performance requirements. The results show that, through all the proposed mechanisms, \approach~significantly outperforms the others with quantification up to 40$\times$ closer to the ground truth while doing so under minimal cognitive effort. To promote open science, all data and code can be accessed at our anonymous repository: \href{https://github.com/ideas-labo/irap}{https://github.com/ideas-labo/irap}.






\section{Related Work}

\textbf{Requirements analytics and quantification.} Requirements automation has evolved from formal methods~\cite{DBLP:journals/re/WhittleSBCB10,DBLP:conf/re/BaresiPS10} to deep semantic analysis via neural language models~\cite{DBLP:conf/re/HeyKKT20,DBLP:conf/kbse/LuoXXS22} and LLMs~\cite{DBLP:journals/corr/abs-2509-13868}. However, these approaches predominantly focus on classification rather than quantification. While \texttt{LQPR}~\cite{DBLP:journals/corr/abs-2511-03421} is a pioneering attempt at rule-based predictive quantification, it permits neither preference reasoning nor interaction at all---the key contributions in this work. 



\textbf{Preference alignment methods based on reinforcement learning.} Methods exist for ensuring the alignment between model outputs and human intentions. Current mainstream paradigms leveraging Reinforcement Learning (RL)~\cite{DBLP:conf/nips/ChristianoLBMLA17,DBLP:conf/nips/Ouyang0JAWMZASR22}, such as DPO~\cite{DBLP:conf/nips/RafailovSMMEF23} and WPO~\cite{DBLP:conf/emnlp/ZhouAZIZSXZ24}, aims to reward the model behaviors that are close to the inputs' needs. A key limitation of those is that they completely leverage LLM to reason the general human preferences (with or without fine-tuning), hence likely to miss the subjective preferences implied in specific cases and cause high uncertainty, especially for performance requirement quantification, which has highly uncertain ambiguity~\cite{DBLP:journals/corr/abs-2511-03421}. Although existing works have attempted to achieve personalized alignment via fine-tuned adapters~\cite{DBLP:conf/iclr/ChengHW24}), maintaining independent models for each user incurs unrealistically excessive costs~\cite{DBLP:journals/corr/abs-2306-07402}.


\textbf{Preference alignment methods based on RAG.} Other methods leverage RAG and LLM in-context learning to achieve dynamic alignment. Often, they iteratively instruct LLM with retrievals to effectively guide model behaviors~\cite{DBLP:conf/icml/BorgeaudMHCRM0L22,DBLP:conf/nips/LewisPPPKGKLYR020}. Nevertheless, existing RAG methods again rely on the implicit instruction of LLMs to integrate bounded context information, without explicitly modeling problem-specific logic of reasoning preferences---this is detrimental for handling uncertain ambiguity in quantifying performance requirements. 





\section{Theory}

\subsection{Quantifying Performance Requirements}
\label{Theory:Qt_form}

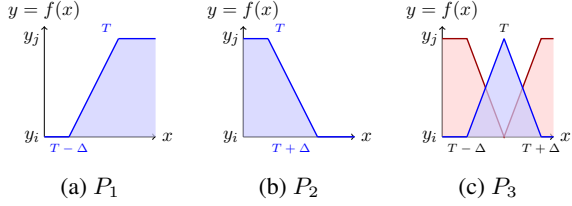
\begin{figure}[t!] 
    \centering
    \begin{subfigure}[b]{0.32\columnwidth} 
        \centering
        \resizebox{\linewidth}{!}{
            \begin{tikzpicture}[scale=0.5]
                \draw[->] (0,0) -- (4.5,0) node[right] {$x$};
                \draw[->] (0,0) -- (0,4.5) node[above] {$y=f(x)$};
                
                \coordinate (A) at (1, 0);
                \coordinate (B) at (3, 4);
                
                \fill[blue!40, opacity=0.7] (0, 0) -- (A); 
                \fill[blue!20, opacity=0.7] (A) -- (B) -- (4.5, 4) -- (4.5, 0) -- (A);
                
                \draw[thick, blue] (0, 0) -- (A) node[below] {\scriptsize $T-\Delta$};
                \draw[thick, blue] (A) -- (B) node[above left] {\scriptsize $T$}; 
                \draw[thick, blue] (B) -- (4.5, 4);
                
                \node at (-0.5, 4) {$y_j$};
                \node at (-0.5, 0) {$y_i$};
            \end{tikzpicture}
        }
        \subcaption{$P_1$} 
        \label{fig:class_lower}
    \end{subfigure}
    \hfill 
    \begin{subfigure}[b]{0.32\columnwidth}
        \centering
        \resizebox{\linewidth}{!}{
            \begin{tikzpicture}[scale=0.5]
                \draw[->] (0,0) -- (4.5,0) node[right] {$x$};
                \draw[->] (0,0) -- (0,4.5) node[above] {$y=f(x)$};
                
                \coordinate (C) at (1, 4);
                \coordinate (D) at (3, 0);
                
                \fill[blue!20, opacity=0.7] (0, 0) -- (0, 4) -- (C) -- (D) -- cycle; 
                \fill[blue!40, opacity=0.7] (D) -- (4.5, 0); 
                
                \draw[thick, blue] (0, 4) -- (C) node[above right] {\scriptsize $T$};
                \draw[thick, blue] (C) -- (D) node[below left] {\scriptsize $T+\Delta$};
                \draw[thick, blue] (D) -- (4.5, 0);
                
                \node at (-0.5, 4) {$y_j$};
                \node at (-0.5, 0) {$y_i$};
            \end{tikzpicture}
        }
        \subcaption{$P_2$} 
        \label{fig:class_upper}
    \end{subfigure}
    \hfill 
    \begin{subfigure}[b]{0.32\columnwidth}
        \centering
        \resizebox{\linewidth}{!}{
            \begin{tikzpicture}[scale=0.5]
                \draw[->] (0,0) -- (4.5,0) node[right] {$x$};
                \draw[->] (0,0) -- (0,4.5) node[above] {$y=f(x)$};
                
                \coordinate (Start) at (0, 4);    
                \coordinate (E) at (1, 0);        
                \coordinate (E_top) at (1, 4);    
                \coordinate (F) at (2.5, 4);      
                \coordinate (F_bot) at (2.5, 0);  
                \coordinate (G) at (4, 0);        
                \coordinate (G_top) at (4, 4);    
                \coordinate (End) at (4.5, 4);    
                
                \fill[red!20, opacity=0.6] (0,0) -- (0,4) -- (1,4) -- (2.5,0) -- (4,4) -- (4.5,4) -- (4.5,0) -- cycle;
                \draw[thick, red!60!black] (0,4) -- (1,4) -- (2.5,0) -- (4,4) -- (4.5,4);

                \fill[blue!20, opacity=0.7] (E) -- (F) -- (G) -- cycle;
                \draw[thick, blue] (0, 0) -- (E) node[below, black] {\scriptsize $T-\Delta$};
                \draw[thick, blue] (E) -- (F) node[above, black] {\scriptsize $T$};
                \draw[thick, blue] (F) -- (G) node[below, black] {\scriptsize $T+\Delta$};
                \draw[thick, blue] (G) -- (4.5, 0); 
                
                \node at (-0.5, 4) {$y_j$};
                \node at (-0.5, 0) {$y_i$};
            \end{tikzpicture}
        }
        \subcaption{$P_3$}
        \label{fig:class_exact}
    \end{subfigure}
    
    \caption{Patterns of the quantification functions for performance requirements.}
    \label{fig:cls}
    \vspace{-0.3cm} 
    
\end{figure}

Inspiring from a recent study~\cite{DBLP:journals/corr/abs-2511-03421}, we found that the requirement on a performance metric exhibits three linear patterns of satisfaction ($y\in[0,1]$), as shown in Figure~\ref{fig:cls}:

\begin{itemize}
    \item \textbf{$P_1$:} Higher value ($x$) up to a threshold $T$ is the most preferred ($y$), and there is a certain tolerance if the value is lower than the threshold (e.g., \textit{``Throughput needs to support above 100 req/s''}).
    \item \textbf{$P_2$:} Lower value down to a threshold $T$ is the most preferred, and there is certain tolerance if the value is higher than the threshold  (e.g., \textit{``Response time is less than 5s''}). 
    \item \textbf{$P_3$:} An exact value is the most preferred/non-preferred; anything lower/higher has some tolerance (e.g., \textit{``Refresh rate shall be equivalent to 5s/time''}).
\end{itemize}

As such, a performance requirement can be quantified as a piecewise function following one (or more) pattern from the above, centered by a threshold/value. Here, the y-axis represents the stratification of stakeholders across the permissive values of a performance metric at the x-axis. For example, consider the natural requirement statement:
\begin{displayquote}
``\textit{The software must receive and process ECG signal data at a frequency of no less than 1000Hz.}''
\end{displayquote}
Clearly, this requirement is vague, as it is not clear to what extent the data frequency less than 1000Hz is tolerable, nor the preference of those that exceed 1000Hz. One possible natural interpretation can be ``\textit{anything better than 1000Hz is equally preferred while there is a tolerance of 10\%, after which it is unacceptable}''. Therefore, the piecewise function of quantification $f(x)$ can be written as: 
\begin{equation}
y=f(x) = 
\begin{cases}
0 & \text{if } x \leq 900 \\
{x-900\over100} & \text{if } 900 < x < 1000 \\
1 & \text{if } x \geq 1000
\end{cases}
\end{equation}
The above follows $P_1$ ($T=1000,\Delta=10\%\times T$). Note that it is possible to have a performance requirement that exhibits multiple patterns\footnote{The number of patterns is often consistent with the number of thresholds~\cite{DBLP:journals/corr/abs-2511-03421}.}, which are joint with several thresholds (and $\Delta$) too (but with their satisfactions adjusted accordingly).

\subsection{Preferences in Quantification}
\label{Theory:prefer}
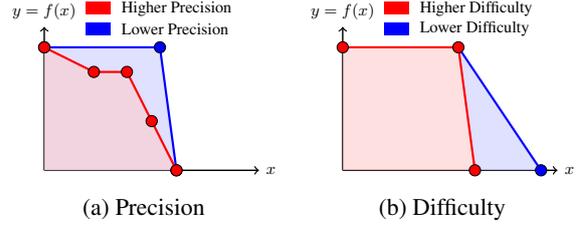
\begin{figure}[t!]
    \centering

    \begin{subfigure}[b]{0.49\columnwidth}
        \centering
        \resizebox{\linewidth}{!}{
            \begin{tikzpicture}[scale=0.8, font=\small] 
                \coordinate (Origin) at (0,0);

                \coordinate (LP1) at (0, 3);
                \coordinate (LP2) at (2.8, 3);
                \coordinate (LP3) at (3.2, 0);

                \coordinate (HP1) at (0, 3);
                \coordinate (HP2) at (1.2, 2.4); 
                \coordinate (HP3) at (2.0, 2.4); 
                \coordinate (HP4) at (2.6, 1.2); 
                \coordinate (HP5) at (3.2, 0);   

                \draw[->, thick] (0,0) -- (5.2,0) node[right] {$x$};
                \draw[->, thick] (0,0) -- (0,3.5) node[above] {$y=f(x)$};

                \fill[blue!20, opacity=0.6] (Origin) -- (LP1) -- (LP2) -- (LP3) -- cycle;
                \fill[red!20, opacity=0.6] (Origin) -- (HP1) -- (HP2) -- (HP3) -- (HP4) -- (HP5) -- cycle;

                \draw[blue, very thick] (LP1) -- (LP2) -- (LP3);
                \foreach \p in {LP1, LP2, LP3} \node[fill=blue, circle,draw=black,scale=0.6] at (\p) {};

                \draw[red, very thick] (HP1) -- (HP2) -- (HP3) -- (HP4) -- (HP5);
                \foreach \p in {HP1, HP2, HP3, HP4, HP5} \node[fill=red, circle,draw=black,scale=0.6] at (\p) {};

                \begin{scope}[shift={(1, 3.8)}]

                    \fill[red] (0, 0) rectangle (0.6, 0.3);
                    \node[right, align=left] at (0.7, 0.15) {\footnotesize {Higher Precision}};

                    \fill[blue] (0, -0.5) rectangle (0.6, -0.2);
                    \node[right, align=left] at (0.7, -0.35) {\footnotesize {Lower Precision}};
                \end{scope}

            \end{tikzpicture}
        }
        \subcaption{Precision}
        \label{fig:precision}
    \end{subfigure}
    \hfill
      \begin{subfigure}[b]{0.49\columnwidth}
        \centering
        \resizebox{\linewidth}{!}{
            \begin{tikzpicture}[scale=0.8, font=\small] 
                \coordinate (Start) at (0, 3);
                \coordinate (Elbow) at (2.8, 3);
                \coordinate (EndOrange) at (3.2, 0);
                \coordinate (EndBlue) at (4.8, 0);
                \coordinate (Origin) at (0,0);

                \draw[->, thick] (0,0) -- (5.2,0) node[right] {$x$};
                \draw[->, thick] (0,0) -- (0,3.5) node[above] {$y=f(x)$};

                \fill[red!20, opacity=0.6] (Origin) -- (Start) -- (Elbow) -- (EndOrange) -- cycle;
                \fill[blue!20, opacity=0.6] (Elbow) -- (EndBlue) -- (EndOrange) -- cycle;

                \draw[red, very thick] (Start) -- (Elbow) -- (EndOrange);
                \draw[blue, very thick] (Elbow) -- (EndBlue);

                \foreach \p/\c in {Start/red, Elbow/red, EndOrange/red, EndBlue/blue}
                    \node[fill=\c, circle,draw=black,scale=0.6] at (\p) {};

                \begin{scope}[shift={(1, 3.8)}]

                    \fill[red] (0, 0) rectangle (0.6, 0.3);
                    \node[right, align=left] at (0.7, 0.15) {\footnotesize {Higher Difficulty}};

                    \fill[blue] (0, -0.5) rectangle (0.6, -0.2);
                    \node[right, align=left] at (0.7, -0.35) {\footnotesize {Lower Difficulty}};
                \end{scope}

            \end{tikzpicture}
        }
        \subcaption{Difficulty}
        \label{fig:difficulty}
    \end{subfigure}

  \caption{Examples of differences on precision and difficulty on performance requirements.}
    \label{fig:prefer}
    \vspace{-0.3cm}
\end{figure}

Indeed, a performance requirement can be interpreted in different ways, and hence its quantification differs. This is highly dependent on what we call preferences of the stakeholders in the requirements, which can be expressed in two aspects:

\begin{itemize}

    \item \textbf{Precision:} This reflects the number of pattern types (and their thresholds) that can be inferred from a performance requirement. Intuitively, a requirement with more patterns leads to a more complicated piecewise function in the quantification, and hence it can be quantified more precisely. With the same example, if there is an additional threshold, e.g., \textit{``...should ideally be 1500Hz''}, then the quantification would have four pieces. An illustration has been shown in Figure~\ref{fig:precision}.

    \item \textbf{Difficulty:} This refers to the value(s) of $T$, $\Delta$, and/or satisfaction from the pattern type(s) when quantifying the requirements. Using the same example before, $\Delta=10\%\times T$ is clearly harder to satisfy than setting $\Delta=50\%\times T$, as illustrated in Figure~\ref{fig:difficulty}.

\end{itemize}

It is worth noting that, commonly, even the stakeholders themselves might not have a clear understanding of their own preferences at the beginning, since without observing the actual quantification, the exact meaning of the performance requirement statements remains cognitively fuzzy in one's mind.

\subsection{Problem Formalization}
\label{Theory:problem_form}

Given the quantification pattern types and the meaning of preferences therein, our goal is to automatically elicit a stakeholder's truly preferred quantification of a performance requirement in an interactive manner. To that end, for quantifying a given target performance requirement, we formulate this as a finite state transition problem:
\begin{equation}
f_{t,0} \xrightarrow{op_1} f_{t,1} \xrightarrow{op_2} \ldots \xrightarrow{op_n} f_{t}^*
\end{equation}
whereby $f_{t,0}$ and $f_t^*$ are the initial and the truly preferred quantification (state) of the performance requirement, respectively; all quantifications between those are intermediate ones. $op_n$ denotes the $n$th operation that includes \textbf{\textsc{add}}, \textbf{\textsc{remove}} pattern types (for controlling precision), and \textbf{\textsc{change}} the values, i.e., $T$, $\Delta$, and the possible satisfactions $y_i$/$y_j$ (for controlling difficulty). With this, our aim is to find the quantification that the stakeholder fully agrees on as soon as possible, i.e., reaching $f_t^*$ with the smallest possible $n$.

\section{\approach~Framework}

\begin{figure}[t!] 
    \centering
    \includegraphics[width=0.85\linewidth]{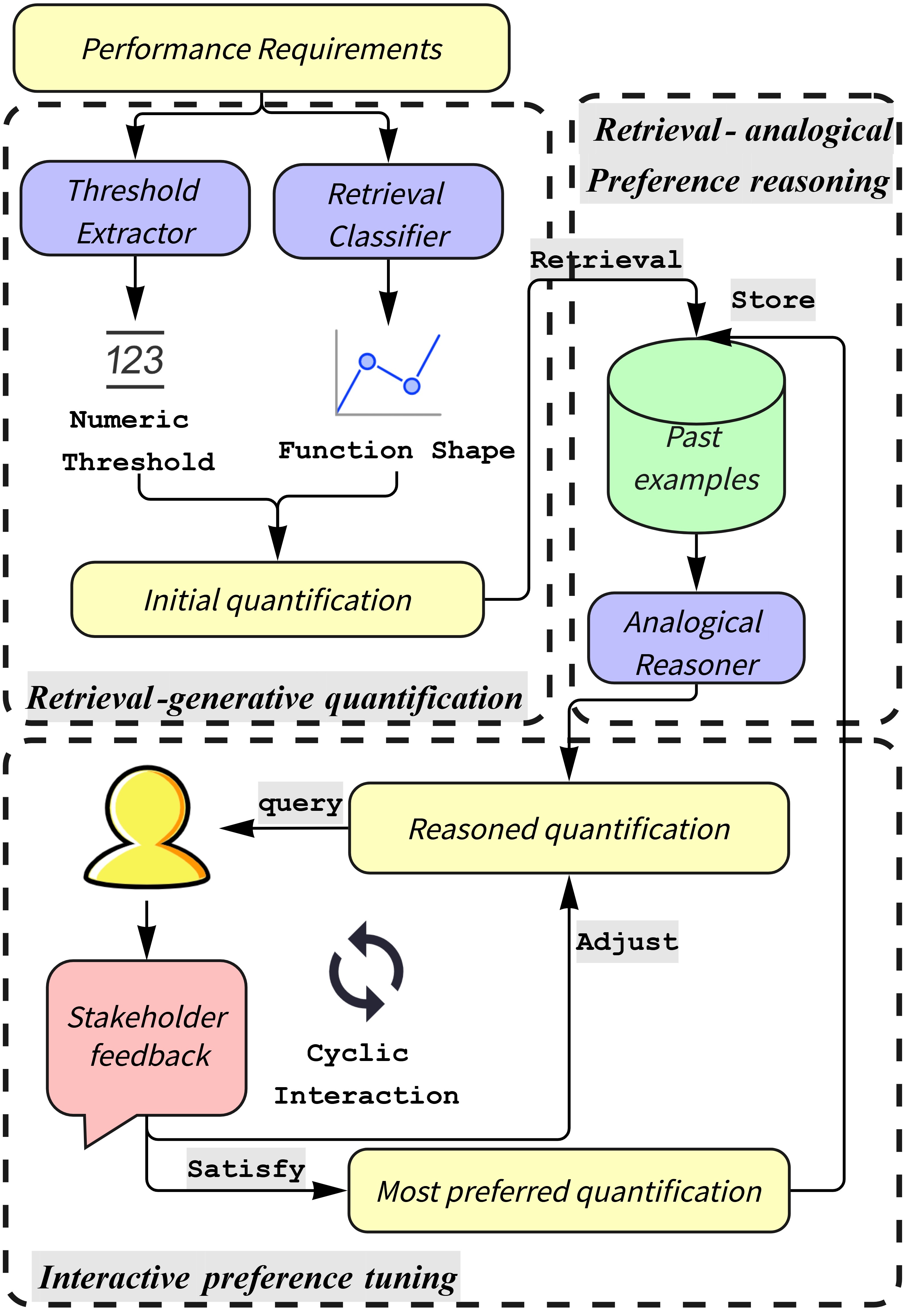}
    \caption{The workflow of \approach.}
    \label{fig:workflow}
\end{figure}

The key property in \approach~is that, with the support of problem-specific knowledge, it not only fosters retrieval reasoning-guided interactive quantification for performance requirements in natural language, but does so with minimized cognitive overhead of the stakeholder, finding a preferred quantification quicker. As in Figure~\ref{fig:workflow}, it has three interrelated phases: (1) A retrieval-generative quantification that converts a given natural requirement statement into the mathematical function of an initial draft quantification; (2) a retrieval-analogical preference reasoning that extracts the past examples as analogies to align the initial quantification closer to what is most likely to be preferred by a stakeholder; (3) mining and tuning preferences via interaction with the stakeholder.

\subsection{Retrieval-Generative Quantification}
    
    From Section~\ref{Theory:Qt_form}, we can formalize the patterns in the forms of points:
    \begin{itemize}
        \item $P_1$: $[(T-\Delta, y_i), (T, y_{j})]$
        \item $P_2$: $[(T, y_i), (T+\Delta, y_{j})]$
        \item $P_3$: $[(T-\Delta, y_i), (T, y_{j}), (T+\Delta, y_i)]$
    \end{itemize}
    Although a requirement might contain more than one pattern, in practice, they can be decomposed into different fragments where each is an independent requirement and involves exactly one such pattern/threshold~\cite{DBLP:journals/corr/abs-2511-03421}. Thus, we model the first phase of quantification as dual tasks: (1) firstly classify the requirement into one pattern via retrieval-based classification; and (2) extract the threshold value using LLM generation. Note that, unlike the threshold $T$, $\Delta$ (default to $\Delta=10\%\times T$) is often implicit and hence can only be adjusted later. This would lead to an initial draft quantification function $f_{t,0}$ for the target.
    
    For example, the performance requirement \textit{``The recommendation accuracy should not be less than 85\%''} can be classified as $P_1$: $[(85-\Delta, 0), (85, 1)]$ where $T=85$.

\subsubsection{Retrieval-based Classification}

Unlike classic natural language classification, where the class label is formulated as independent one-hot indices without any of their semantic information, \approach~embeds the semantic knowledge of the label into the classification, which has been shown to be superior~\cite{DBLP:conf/acl/HenaoLCSWWZZ18}. This is because, as studied before~\cite{DBLP:journals/corr/abs-2511-03421}, performance requirements exhibit clear phrases that can pinpoint their patterns, serving as a solid foundation for incorporating the semantics of labels. Specifically, we firstly extract 10 key phrases from known performance requirements for each pattern from Section~\ref{Theory:Qt_form}, as shown in Table~\ref{tab:anchor_examples}, representing their respective semantic information. Those phrases serve as the anchors.

Secondly, since those anchors might not be exhaustive, we seek to extract the semantic meaning from a performance requirement and match it with the semantics of each anchor, which outperforms simple syntactical matching. To that end, an embedding model is needed, and we use the RoBERTa~\cite{DBLP:journals/corr/abs-1907-11692}. However, directly using existing fine-tuning loss is ill-suited, because we found that standard fine-tuning struggles to distinguish between antonymous anchors like \textit{``at least''} ($P_1$) and \textit{``at most''} ($P_2$) that appear in identical contexts, hence we need stronger, specifically crafted guidance to discriminate those cases. To resolve that, we fine-tune the RoBERTa using a contrastive loss extended from the \texttt{InfoNCE} loss~\cite{DBLP:journals/corr/abs-1807-03748}, aiming to globally maximize the average similarity between the input embedding of requirements/anchors and all their matching patterns while penalizing the similarity with all non-matching ones, bridging the requirements and anchors. The loss function is:
\begin{equation}
\small
\mathcal{L}(\mathcal{S}) = \sum_{s_i \in \mathcal{S}}{ \frac{1}{|\mathcal{P}_i|} \sum_{p \in \mathcal{P}_i} \left( - \log \frac{\exp(\text{sim}(s_i, p)/\tau)}{\sum_{a \in \mathcal{A}} \exp(\text{sim}(s_i, a)/\tau)} \right) }
\end{equation}
where $\text{sim}(u, v)$ denotes cosine similarity, $\tau$ is the temperature parameter. $s_i$ denotes a performance requirement/anchor. $\mathcal{P}_i$ is the set of matching patterns for $s_i$, and $\mathcal{A}$ denotes all patterns. In this way, the obtained semantics of both the requirement statement and the anchor phrases would contain strong semantics of the labeled pattern class.

Finally, we calculate the cosine similarity between the embedding of the given performance requirement and each of the anchors, assigning the pattern of the anchor with the highest similarity.

    
    

\begin{table}[t!]
\centering
\adjustbox{max width=\columnwidth}{
\begin{tabular}{ll}
 \toprule
\textbf{Pattern Type} & \textbf{Anchor Phrases} \\ 
 \midrule
\textbf{$P_1$}  & \textit{``no less than''}, \textit{``at least''}, \textit{``greater than''}, $\dots$ \\ 
\textbf{$P_2$} & \textit{``no more than''}, \textit{``at most''}, \textit{``less than''}, $\dots$ \\ 
\textbf{$P_3$}  & \textit{``exactly''}, \textit{``precisely''}, \textit{``equivalent to''}, $\dots$ \\ 
 \bottomrule
\end{tabular}
}
\caption{Exampled anchors; full list is at Appendix~\ref{append:anchor_phrases}.}
\label{tab:anchor_examples}
\end{table}

\subsubsection{Generative Threshold Extraction}

While we are able to classify a performance requirement statement into patterns, the threshold still needs to be identified. To extract the threshold, \approach~achieves such via fine-tuning a LLM with full parameters, i.e., GPT-2 (774M) in this case, as not all numbers in the requirement statement are thresholds, and we need to have a comprehensive understanding of the statement. Notably, GPT-2 is chosen since we do not want an over-complex heavy model, and GPT-2 is lightweight, efficient, and with sufficient accuracy. The sample/prompt can be found at Appendix~\ref{sec:thres}

\subsection{Retrieval-Analogical Preference Reasoning}
\label{method:stage2}

A naive approach would be to directly use the $f_{t,0}$ from retrieval-generative quantification for the stakeholder to start interacting with. However, this could incur large cognitive overhead as the generative one might still be far away from one's true preference. Since each stakeholder can propose many performance requirements, \approach~seeks to gradually align the subjective preferences of the stakeholder deriving from the previously quantified requirements, hence creating a better starting point of interaction for the newly given requirement statement $s_t$ that can reach $f_{t}^*$ quicker. Here, assuming that the quantification of a requirement can be represented by $z$ points, e.g., $f_{k,0}: \{(x_{k,1}, y_{k,1}), \dots, (x_{k,z}, y_{k,z})\}$, we retrieve the most semantically-similar\footnote{We embed each past requirement and $s_t$ using BERT and compute their cosine similarity.} past requirement $s_k$ to the $s_t$, such that the initial quantification $f_{k,0}$ of $s_k$ has the same number of points as $f_{t,0}$, from those that have been quantified by \approach. Formally, $s_k$ is expressed as: 
\begin{equation}
s_{k}=\{f_{k,0}, f_{k}^*\}   
\end{equation}
whereby $f_{k,0}$ and $f_{k}^*$ denote the initial quantification and the finally accepted, most preferred quantification, respectively. Our goal is to extract the proper operations for converting from $f_{k,0}$ to $f^*_k$ as the analogy, and then apply them to $f_{t,0}$.

Yet, since the set of operations we can use for transferring from one quantification to another is vast especially when they have different pieces/points, finding the operations for converting from $f_{k,0}$ to $f^*_k$ is not straightforward. Suppose that we have $f_{k,0}: \{(9, 0), (10, 1)\}$ and $f^*_{k}:\{(8.5, 0), (9.5, 0.5), (10.5, 1)\}$. Now, from $f_{k,0}$ to $f_{k}^*$ we can either add $(9.5, 0.5)$, change $(9, 0)$ to $(8.5, 0)$, and $(10, 1)$ to $(10.5, 1)$; or add $(8.5, 0)$, change $(9, 0)$ to $(9.5, 0.5)$, and $(10, 1)$ to $(10.5, 1)$. Both are valid, but their distances differ, e.g., using edit distance\footnote{We count adding/ removing the whole point and changing each value as independent operations, e.g., adding $(9, 0)$ is one operation; changing $(9, 0)$ to $(9.5, 0.5)$ would be two operations as there are two value changes.}, the former has 3 while the latter is 4. Clearly, the former is wiser with smaller magnitudes of changes on the quantification function.

\begin{figure}[t!] 
    \centering
    \begin{subfigure}[b]{0.23\linewidth}
        \centering
        \includegraphics[width=\linewidth]{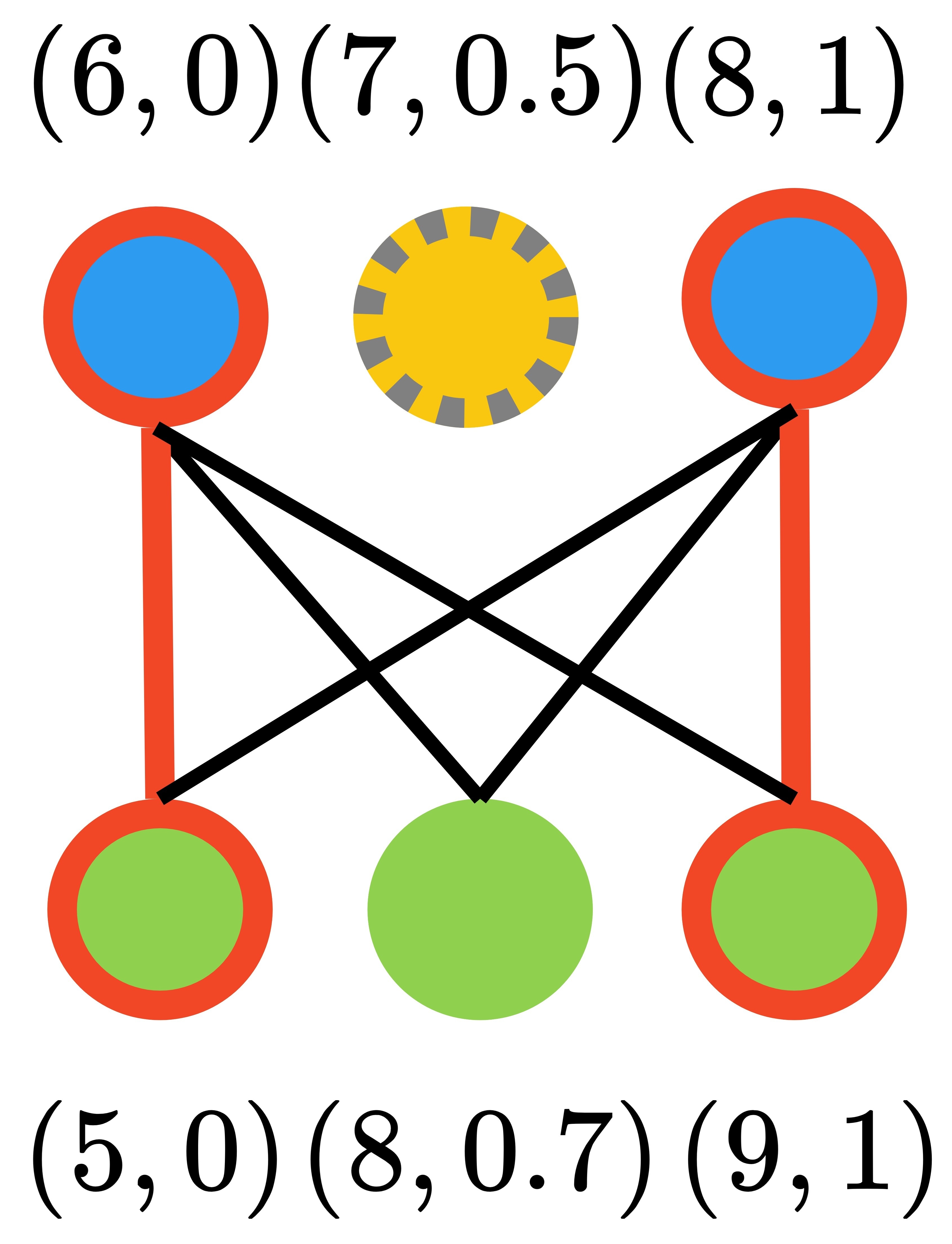}
       \caption{Case 1 (\textsc{add})}
        \label{fig:bg1}
    \end{subfigure}
    \hfill
    \begin{subfigure}[b]{0.23\linewidth}
        \centering
        \includegraphics[width=\linewidth]{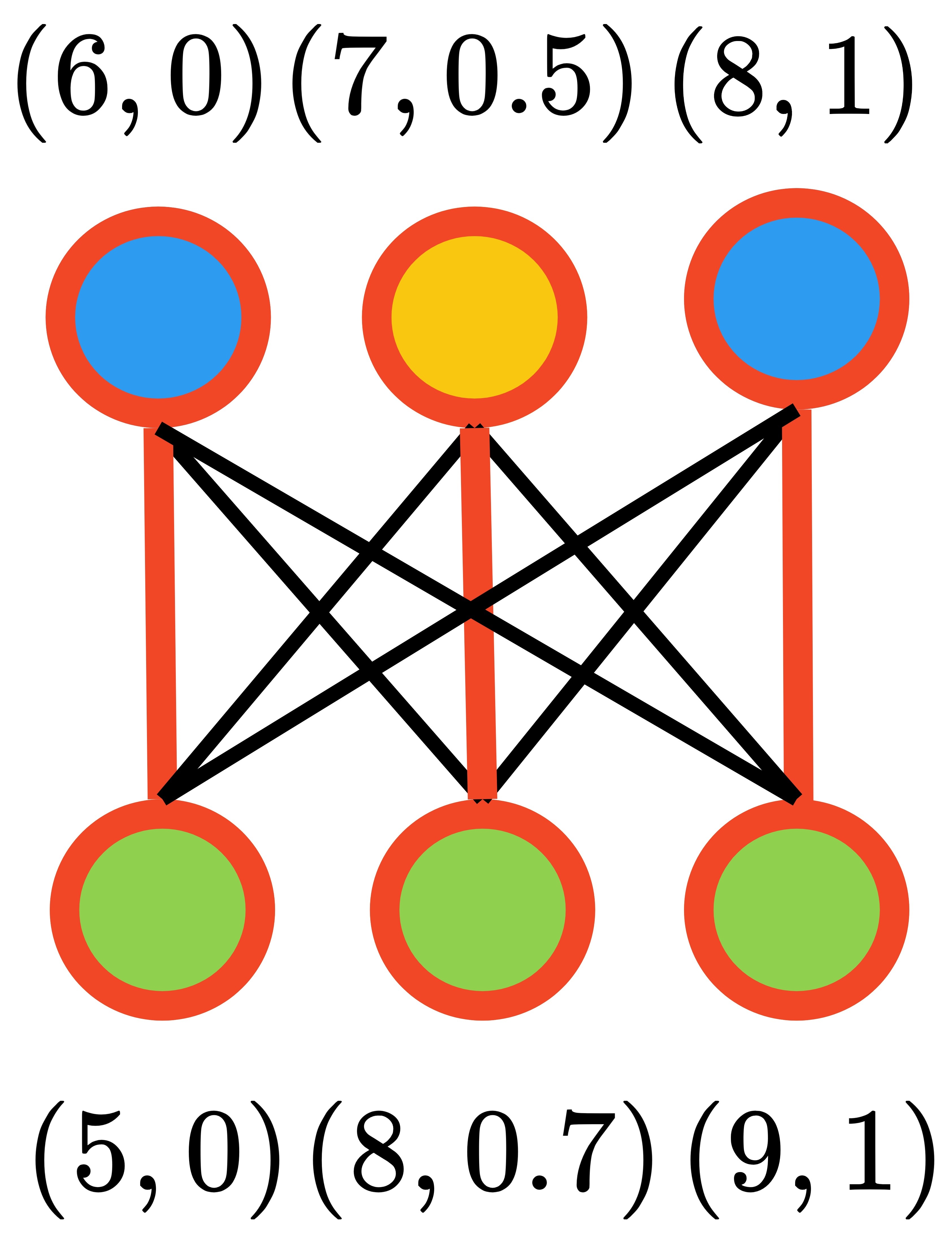}
        \caption{Case 1 (\textsc{change})}
        \label{fig:bg2}
    \end{subfigure}
    \hfill
    \begin{subfigure}[b]{0.23\linewidth}
        \centering
        \includegraphics[width=\linewidth]{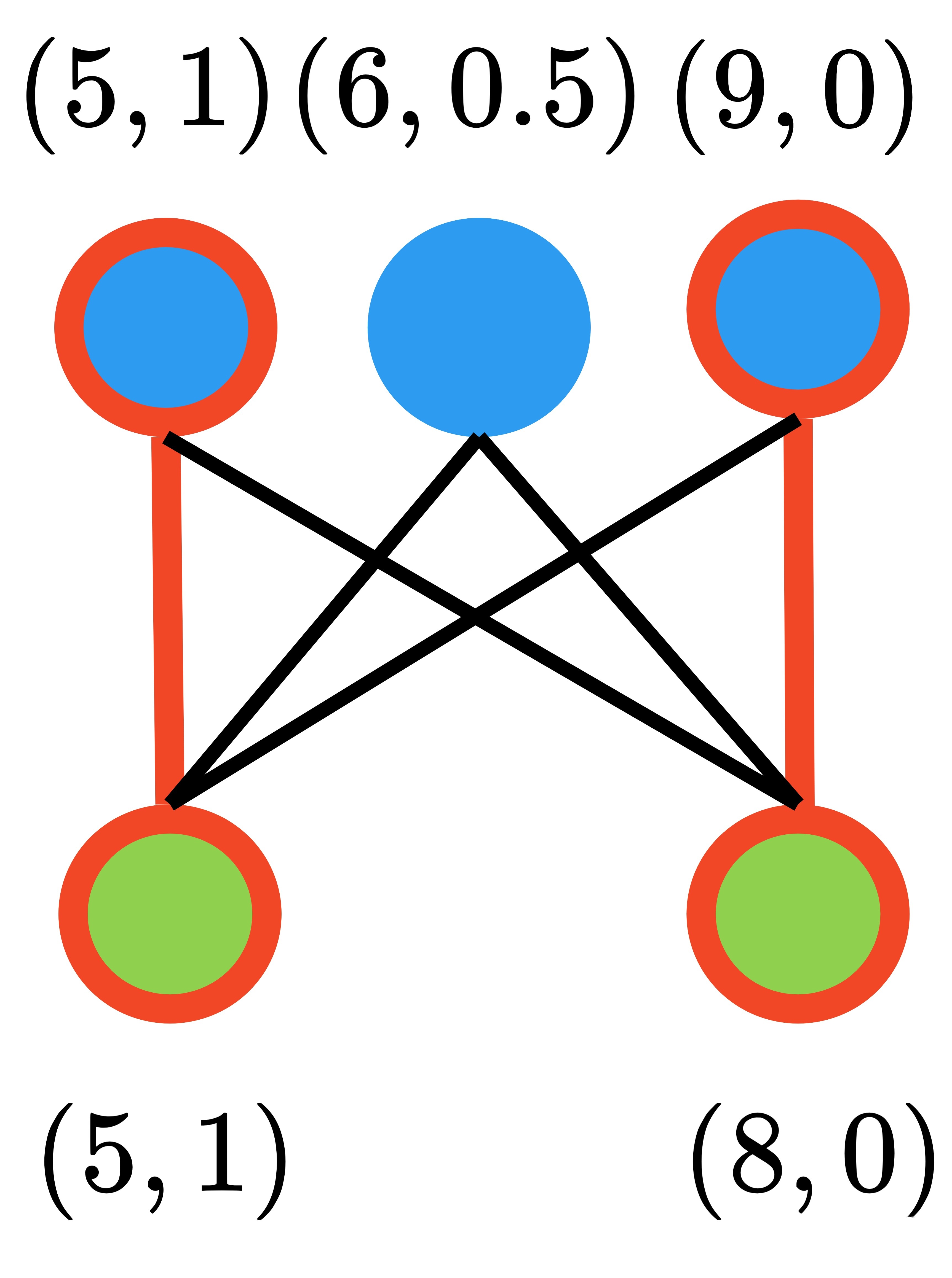}
       \caption{Case 2 (\textsc{remove})}
        \label{fig:bg3}
    \end{subfigure}
    \hfill
    \begin{subfigure}[b]{0.23\linewidth}
        \centering
        \includegraphics[width=\linewidth]{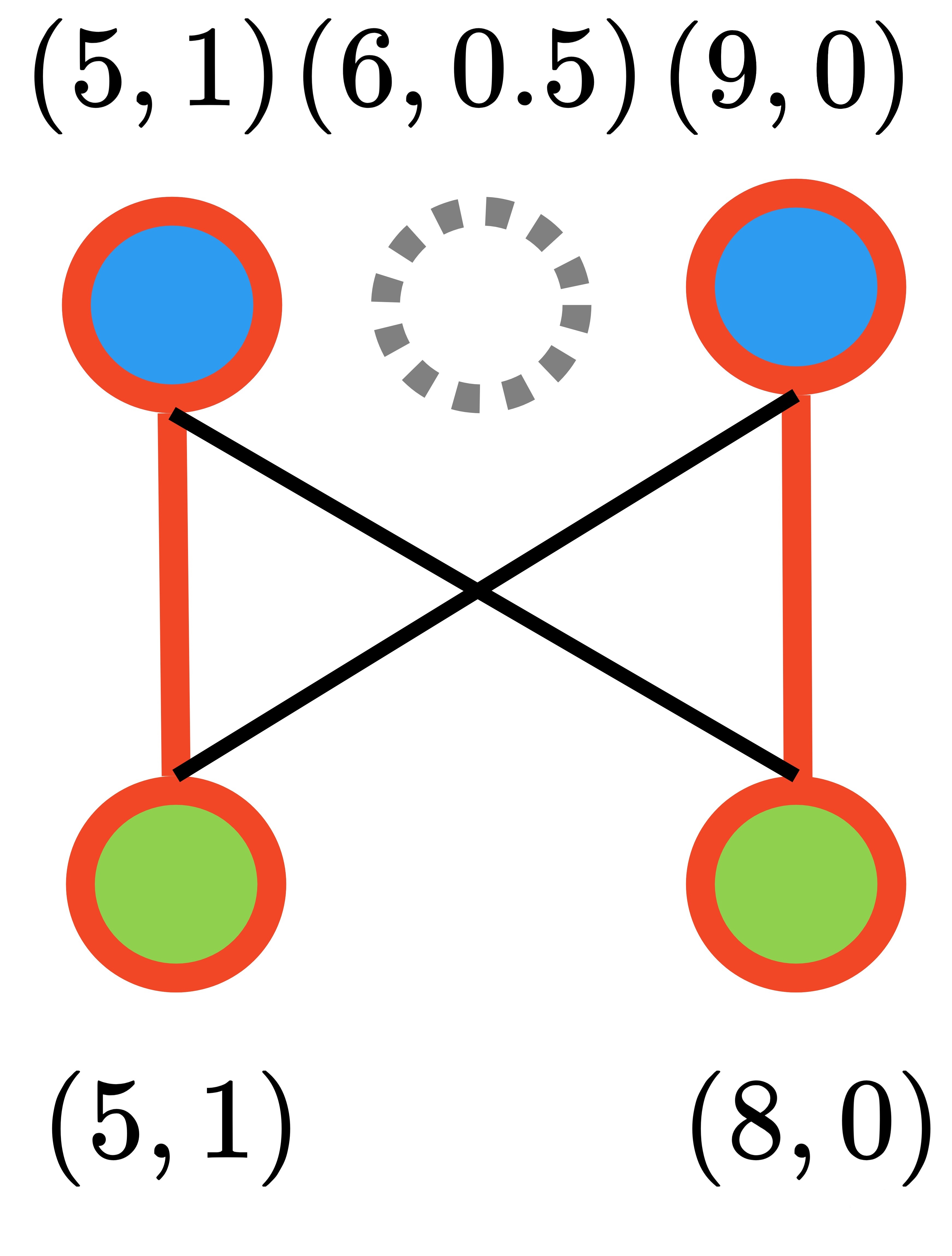}
        \caption{Case 2 (\textsc{change})}
        \label{fig:bg4}
    \end{subfigure}

    \caption{Points alignment (a and c) and changes identification (b and d). Mappings are highlighted.}
    \label{fig:bi_graph}
\end{figure}

To accurately measure the converting distance from $f_{k,0}$ to $f_{k}^*$ and obtain their operations, we propose a path-aware operation extraction (PAOE), such that not only the actual change, but also the weights/costs with the change, are included. We formalize this as a bipartite graph maximum weight matching problem~\cite{DBLP:books/daglib/p/Kuhn10} based on a graph for $f_{k,0}$ and $f_{k}^*$: 
\begin{itemize}
    \item \textbf{Vertices:} As in Figure \ref{fig:bi_graph}, we list the points in $f_{k,0}$ as the top vertex set $\mathcal{U} = \{u_1, u_2, \dots, u_n\}$, and list those in $f_{k}^*$ to the bottom vertex set $\mathcal{V} = \{v_1, v_2, \dots, v_m\}$.
    \item \textbf{Edges and weights:} The set $\mathcal{E}$ contains edges connecting any node in $\mathcal{U}$ with any node in $\mathcal{V}$. The weight $w_{ij}$ of the edge $(u_i, v_j)$ represents the ``matching preference'' between point $u_i$ and $v_j$. We define the edge weight as the negative Euclidean distance: $w_{ij} = - \sqrt{(x_{u_i} - x_{v_j})^2 + (y_{u_i} - y_{v_j})^2}$.
\end{itemize}

We seek to find a matching $\mathcal{M}$ in the graph (i.e., a subset of edges, where any two edges do not share a vertex), such that the sum of the weights of all edges within the matching is maximized:
\begin{equation}
\arg \max_{\mathcal{M}} \sum_{(u_i, v_j) \in \mathcal{M}} w_{ij}
\end{equation}
The above can be solved by the well-known Kuhn-Munkres (KM) algorithm~\cite{DBLP:books/daglib/p/Kuhn10}.





Drawing on the most semantically-similar requirement $s_k$ and PAOE, the key steps of retrieval-analogical preference reasoning in \approach~are:

\begin{enumerate}
    \item \textbf{Points alignment:} We firstly employ the KM algorithm to find the optimal matching between points in $f_{k,0}$ and $f_{k}^*$. If there are more points in $f_{k,0}$ than in $f_{k}^*$, then the unmatched points in $f_{k,0}$ are subject to \textsc{remove} (Figure~\ref{fig:bi_graph}c). When there are more points in $f_{k}^*$ than in $f_{k,0}$, we identify the unmatched points in $f_{k}^*$ as those that require \textsc{add} the corresponding points in $f_{k,0}$ and set their $x$/$y$ by averaging those of the two adjacent points, e.g., $(7,0.5)$ in Figure~\ref{fig:bi_graph}a.
    
    \item \textbf{Changes identification:} The KM algorithm is then run again on the aligned points. The final matching of points with different values (including newly added points) would indicate those that need \textsc{change} on $T$, $\Delta$, and/or the satisfaction $y_i$/$y_j$ (Figures~\ref{fig:bi_graph}b and~\ref{fig:bi_graph}d).

    \item \textbf{Operations sequencing:} {All the operations incurred from the previous two steps serve as the analogy, in which the \textsc{add} and \textsc{remove} are always placed before \textsc{change}\footnote{This is a simple and pragmatic choice to eliminate the chance of having an invalid operation sequence, e.g., we cannot change a value before it is added. Without this, we would need a more expensive strategy to repair the above case, which is not ideal.}.}

\end{enumerate}
As such, the extracted operations can be applied to the $f_{t,0}$ for obtaining a new initial and reasoned quantification $f'_{t,0}$ for the interaction phase. The detailed pseudo code can be found at Appendix~\ref{append:analogical_reasoning}.

Taking Case 2 in Figure~\ref{fig:bi_graph} as an example, the operations are \textsc{remove} (the second point) and \textsc{change} (decease the $x$ of the new second point\footnote{The extent of change is adjusted and bounded by the function, starting with $10\%$ of the original value.}). Suppose that $f_{t,0} : \{(25, 1), (30, 0.5), (40, 0)\}$, then applying the two operations has: 
\begin{flalign*}
& \{(25, 1), (30, 0.5), (40, 0)\} \xrightarrow{op_1} \{(25, 1), (40, 0)\} & \\
& \xrightarrow{op_2} \{(25, 1), (36, 0)\} &
\end{flalign*}
leading to the new $f_{t,0}^{'}:\{(25, 1), (36, 0)\}$.

PAOE is also efficient, e.g., for processing requirement under $40+$ past examples, it takes $<0.2$ seconds on a 2.8GHz and 8GB RAM machine.

\subsection{Interactive Preference Tuning}

With the converted/reasoned $f'_{t,0}$, \approach~can then start iteratively interactive quantification with the stakeholder. Here, we follow tree-based multiple choice questions-answering for preference tuning, since this can significantly reduce stakeholders' cognitive load and improve efficiency~\cite{polat2020analysis}, leading to a most preferred quantification $f_{t}^*$ quicker. A snippet of the questions has been shown in Figure~\ref{fig:part_inqury}. In total, there are 5 levels with 7 (alternative) questions, which can be queried iteratively over rounds (a round runs from the root to one of the leaves). Each leaf denotes an operation related to the precision or difficulty, i.e., \textsc{add} (values are set by averaging adjacent points as before), \textsc{remove}, or \textsc{change} ($T$, $\Delta$, and/or satisfaction by a small step). The chosen operation is applied to the quantification at the end of a round.

Note that if we allow infinite interaction rounds, then certainly \approach~would always reach the most preferred quantification, but it is too costly. As such, we set a hyperparameter $N$ which bounds the maximum number of rounds in the interaction.

\input{figures/query_tree}

\section{Evaluation}

\subsection{Experiment Setup}

\textbf{Datasets and Procedure:} As in Table~\ref{tab:dataset}, we use four carefully curated datasets, extracting the performance requirements documented in real-world projects, as the testing sets. Since real-world samples are limited, we leveraged GPT-4 to generate synthetic data based on examples from 34 software engineering domains~\cite{DBLP:conf/icse/FerraraCGLP24}, leading to $2,560$ (labeled) performance requirements as the initial quantification examples for the fine-tunings and retrieval knowledge in \approach. For each dataset, the ground truth---the most preferred quantification of the requirements---is manually agreed and annotated by the authors and collaborators who are experienced software engineers. This is important for us to conduct fair comparisons and they still represent real-world preferences.

The annotation follows two steps: 

\begin{itemize}
    \item Determine the number of ``break points'' that break the quantification into two extra fragments. This is basically a classification task, and we computed Cohen’s $\kappa$~\cite{landis1977measurement} as an indicator of inter-annotator agreement rates, then iterated the process until there is a $\kappa>0.7$ (a pragmatic threshold of reliable common agreement in prior works).
    \item Determine the value of each break point. For each of those points, we then voted for eliminating one value in turn, until only one final value is left, which would be chosen.
\end{itemize}


For interaction, we ask several experienced human software engineers (five per dataset) to use \approach~with the annotated ground truth in mind. There are 5 repeats for all methods, even if no humans are invovled.





\textbf{State-of-the-art Methods:} We compare \approach~against four categories of 10 state-of-the-art methods: (1) domain-specific method: the rule- and pattern-based \texttt{LQPR}~\cite{DBLP:journals/corr/abs-2511-03421};  (2) the vanilla LLMs, i.e., \texttt{DeepSeek-V3}, \texttt{Qwen3-Coder}, \texttt{gpt-5-mini}, and \texttt{llama-4}; (3) RAG-based methods: Naive \texttt{RAG}~\cite{DBLP:conf/nips/LewisPPPKGKLYR020}, \texttt{Hybrid RAG}~\cite{DBLP:journals/tois/BruchGI24}, and the noise-resilient \texttt{ASTUTE RAG}~\cite{DBLP:conf/acl/WangW00A25}, for all of which \texttt{Qwen3-Coder} (480B) is used as the backbone LLM; (4) the preference-optimized methods with RL: DPO~\cite{DBLP:conf/nips/RafailovSMMEF23} and WPO~\cite{DBLP:conf/emnlp/ZhouAZIZSXZ24}, which are paired with a fine-tuned \texttt{Qwen-7B} via LoRA~\cite{DBLP:conf/iclr/HuSWALWWC22}. As with \approach, the applicable methods use the same synthetic dataset for fine-tuning and knowledge retrieval. The prompt details are at Appendices~\ref{sec:llm}--\ref{sec:rl}. For methods that support direct human interactions on-the-fly (with any questions), e.g., the vanilla LLMs, we ask the same human software engineers and set the same maximum interaction rounds as \approach~($N=5$).


\begin{table}[t!]
\centering

\adjustbox{max width=\columnwidth}{
\begin{tabular}{llll}
 \toprule

\textbf{Dataset} & \textbf{$\#$ Projects} & \textbf{$\#$ Perf. Req.} & \textbf{Source}  \\

\midrule

\textsc{Promise}~\cite{promise}&15&45   &real-world\\
\textsc{PURE}~\cite{DBLP:conf/re/FerrariSG17}&79&23&real-world\\
\textsc{SRS}~\cite{DBLP:conf/cse/ShaukatNZ18}&4&15&real-world\\
\textsc{FQ}~\cite{DBLP:journals/corr/abs-2504-16768}&5&10&real-world\\
LLM-generated dataset&N/A&2560&synthetic\\

\bottomrule
\end{tabular}
}
\caption{Details of the datasets studied.}
\label{tab:dataset}
\end{table}

\begin{table*}[t!]
\centering
\scriptsize
\renewcommand{\arraystretch}{1.1}
\setlength{\tabcolsep}{1.6mm}
\begin{adjustbox}{width=\linewidth, center}
\begin{tabular}{l l c c c c | c c c c}
\toprule
\textbf{Type} & \textbf{Method} & P2P & Chebyshev & RMSE & IAD & P2P & Chebyshev & RMSE & IAD \\
\midrule
& & \multicolumn{4}{c}{\cellcolor{gray!20}\textbf{\textit{\textsc{Promise} Dataset}}} & \multicolumn{4}{c}{\cellcolor{gray!20}\textbf{\textit{\textsc{PURE} Dataset}}} \\
\cmidrule(lr){3-6} \cmidrule(lr){7-10} 
Domain-specific & \texttt{LQPR}~\cite{DBLP:journals/corr/abs-2511-03421} & 0.950 (0.340) & 0.460 (0.280) & 0.300 (0.210) & 0.240 (0.190) & 0.870 (0.380) & 0.570 (0.320) & 0.320 (0.240) & 0.260 (0.210) \\
\hline
\multirow{4}{*}{Vanilla LLMs} & \texttt{DeepSeek-V3} (671B) & 0.774 (0.451) & 0.491 (0.331) & 0.274 (0.231) & 0.201 (0.191) & 0.651 (0.451) & 0.481 (0.391) & 0.251 (0.231) & 0.201 (0.211) \\
& \texttt{Qwen3-Coder}(480B) & 0.851 (0.584) & 0.481 (0.371) & 0.224 (0.191) & 0.141 (0.145) & 0.651 (0.491) & 0.431 (0.341) & 0.231 (0.201) & 0.191 (0.181) \\
& \texttt{gpt-5-mini} (300B) & 0.751 (0.471) & 0.474 (0.281) & 0.264 (0.191) & 0.191 (0.154) & 0.791 (0.311) & 0.391 (0.371) & 0.201 (0.211) & 0.161 (0.191) \\
& \texttt{llama-4} (300B) & 0.864 (0.381) & 0.341 (0.391) & 0.191 (0.231) & 0.141 (0.181) & 0.791 (0.311) & 0.391 (0.371) & 0.201 (0.211) & 0.161 (0.191) \\
\hline
\multirow{3}{2.2cm}{RAG-based (\texttt{Qwen3-Coder} 480B)}  & \texttt{RAG}~\cite{DBLP:conf/nips/LewisPPPKGKLYR020} & 0.549 (0.536) & 0.310 (0.394) & 0.178 (0.247) & 0.140 (0.202) & \textbf{0.326 (0.559)} & \textbf{0.217 (0.355)} & 0.126 (0.208) & 0.103 (0.178) \\
& \texttt{Hybrid RAG}~\cite{DBLP:journals/tois/BruchGI24} & 0.699 (0.542) & 0.349 (0.404) & 0.178 (0.228) & 0.125 (0.175) & 0.949 (0.511) & 0.700 (0.379) & 0.418 (0.278) & 0.300 (0.268) \\
& \texttt{ASTUTE RAG}~\cite{DBLP:conf/acl/WangW00A25} & 0.760 (0.569) & 0.451 (0.339) & 0.227 (0.182) & 0.147 (0.140) & 0.617 (0.369) & 0.463 (0.267) & 0.251 (0.169) & 0.185 (0.162) \\
\hline
\multirow{2}{2.2cm}{Preference-optimized with RL (\texttt{Qwen-7B})} & \texttt{DPO}~\cite{DBLP:conf/nips/RafailovSMMEF23} & 1.252 (0.866) & 0.774 (0.328) & 0.480 (0.277) & 0.338 (0.287) & 1.356 (1.292) & 0.870 (0.221) & 0.546 (0.208) & 0.374 (0.265) \\
& \texttt{WPO}~\cite{DBLP:conf/emnlp/ZhouAZIZSXZ24} & 1.470 (0.753) & 0.931 (0.205) & 0.439 (0.318) & 0.323 (0.277) & 1.672 (0.918) & 0.992 (0.039) & 0.605 (0.170) & 0.428 (0.164) \\
\hline
& \approach~(Ours) & \textbf{0.239 (0.385)} & \textbf{0.029 (0.105)} & \textbf{0.014 (0.052)} & \textbf{0.010 (0.042)} & \textbf{0.041 (0.054)} & \textbf{0.136 (0.218)} & \textbf{0.055 (0.085)} & \textbf{0.034 (0.050)} \\
\bottomrule
\toprule
& & \multicolumn{4}{c}{\cellcolor{gray!20}\textbf{\textsc{SRS} \textit{Dataset}}} & \multicolumn{4}{c}{\cellcolor{gray!20}\textbf{\textit{\textsc{FQ} Dataset}}} \\
\cmidrule(lr){3-6} \cmidrule(lr){7-10} 
Domain-specific & \texttt{LQPR}~\cite{DBLP:journals/corr/abs-2511-03421} & 0.740 (0.310) & 0.400 (0.260) & 0.160 (0.180) & 0.210 (0.150) & 0.880 (0.350) & 0.760 (0.290) & 0.510 (0.220) & 0.340 (0.190) \\
\hline
\multirow{4}{*}{Vanilla LLMs} & \texttt{DeepSeek-V3} (671B) & 0.621 (0.381) & 0.401 (0.331) & 0.221 (0.211) & 0.161 (0.171) & 0.854 (0.461) & 0.642 (0.342) & 0.322 (0.185) & 0.241 (0.192) \\
& \texttt{Qwen3-Coder} (480B) & 0.861 (0.291) & 0.551 (0.271) & 0.291 (0.221) & 0.224 (0.201) & 1.084 (0.742) & 0.741 (0.371) & 0.451 (0.271) & 0.371 (0.268) \\
& \texttt{gpt-5-mini} (300B) & 0.761 (0.291) & 0.531 (0.281) & 0.241 (0.141) & 0.161 (0.125) & 0.932 (0.401) & 0.651 (0.351) & 0.374 (0.242) & 0.301 (0.265) \\
& \texttt{llama-4} (400B) & 0.691 (0.231) & 0.281 (0.311) & 0.131 (0.161) & 0.098 (0.125) & 0.801 (0.374) & 0.684 (0.351) & 0.391 (0.201) & 0.322 (0.171) \\
\hline
\multirow{3}{2.2cm}{RAG-based (\texttt{Qwen3-Coder} 480B)}   & \texttt{RAG}~\cite{DBLP:conf/nips/LewisPPPKGKLYR020} & 0.415 (0.600) & 0.218 (0.355) & 0.114 (0.187) & 0.081 (0.140) & 0.286 (0.410) & 0.026 (0.132) & 0.014 (0.070) & 0.013 (0.063) \\
& \texttt{Hybrid RAG}~\cite{DBLP:journals/tois/BruchGI24} & 1.092 (0.580) & 0.760 (0.336) & 0.401 (0.219) & 0.275 (0.191) & 0.332 (0.470) & 0.093 (0.260) & 0.059 (0.168) & 0.046 (0.143) \\
& \texttt{ASTUTE RAG}~\cite{DBLP:conf/acl/WangW00A25} & 0.689 (0.569) & 0.395 (0.278) & 0.202 (0.167) & 0.138 (0.148) & 0.747 (0.677) & 0.516 (0.417) & 0.276 (0.242) & 0.188 (0.212) \\
\hline
\multirow{2}{2.2cm}{Preference-optimized with RL (\texttt{Qwen-7B})}  & \texttt{DPO}~\cite{DBLP:conf/nips/RafailovSMMEF23} & 1.342 (1.353) & 0.878 (0.257) & 0.542 (0.237) & 0.375 (0.284) & 1.373 (0.821) & 0.831 (0.271) & 0.526 (0.256) & 0.418 (0.308) \\
& \texttt{WPO}~\cite{DBLP:conf/emnlp/ZhouAZIZSXZ24} & 1.903 (1.280) & 0.981 (0.052) & 0.645 (0.124) & 0.539 (0.148) & 0.956 (0.408) & 0.973 (0.044) & 0.532 (0.213) & 0.355 (0.142) \\
\hline
& \approach~(Ours) & \textbf{0.100 (0.221)} & \textbf{0.130 (0.238)} & \textbf{0.051 (0.094)} & \textbf{0.030 (0.056)} & \textbf{0.270 (0.412)} & \textbf{0.000 (0.000)} & \textbf{0.000 (0.000)} & \textbf{0.000 (0.000)} \\
\bottomrule
\end{tabular}
\end{adjustbox}
\caption{Mean (deviation) performance over all repeats and requirements. \textbf{Bold} highlights the best result per case.}
\label{tab:main_results}
\end{table*}

\textbf{Evaluation Metrics:} We use several metrics to assess the difference between the produced and the ground truth quantification for each test requirement: we use Point-to-Point Distance (P2P) to assesses the structure similarity~\cite{DBLP:books/daglib/p/Kuhn10} and Maximum Deviation (Chebyshev) that quantifies the worst-case vertical discrepancy~\cite{powell1981approximation} as two \textit{geometric} metrics; we also use two \textit{scalar} metrics Root Mean Square Error (RMSE) that measures the average deviation of $y$ values~\cite{chai2014root}; and Integrated Area Difference (IAD) that evaluates the difference of requirement difficulty by area~\cite{DBLP:journals/ijcga/AltG95}. For all metrics, the smaller the value, the better. More details can be found at Appendix~\ref{sec:app_m}.

\subsection{Main Results}

Table \ref{tab:main_results} presents the results over all repeats and requirements. As can be seen, in general, \approach~performs remarkably better than the others, ranking the first for all datasets/metrics with up to $40\times$ improvements. Rule-based methods like \texttt{LQPR} can only quantify performance requirements as patterns and lack preference reasoning capabilities. Directly instructing LLMs (e.g., \texttt{DeepSeek-V3}) performs poorly too, since it remains challenging to ensure that the LLM can fully understand the complexity of requirement quantification even with interactions. Preference-optimized methods with RL, such as \texttt{DPO} are also not performing well, since they fail to capture the imprecise and vague preference implied in a performance requirement. The RAG-based methods perform the best among the state-of-the-art methods, since the retrieval is important for quantification. However, they remain inferior to \approach~because they do not explicitly handle the problem-specific patterns/knowledge as \approach. A qualitative case study can be found at Appendix~\ref{sec:add_quanlitative}.

\subsection{Ablation Study}
\input{figures/ablation}

We respectively remove the retrieval-analogical preference reasoning (\texttt{w/o-A}), interactive preference tuning (\texttt{w/o-T}), and both (\texttt{w/o-AT}) from \approach~for comparisons. As we can see from Figure~\ref{fig:ablation_visual}, \approach~indeed generally performs the best compared with the others. Notably, \texttt{w/o-AT} performs much worse than \texttt{w/o-A} and \texttt{w/o-T}, meaning that while both retrieval-analogical preference reasoning and interactive preference tuning make non-trivial contributions to \approach~individually, their combination is the key to its success.

\subsection{Cognitive Overhead}
\begin{table}[t!]
\centering
\scriptsize
\setlength{\tabcolsep}{2pt} 

\begin{tabularx}{\columnwidth}{l @{\extracolsep{\fill}} c c c c}
\toprule
\textbf{Dataset} & \textbf{\texttt{DeepSeek-V3}} & \textbf{\texttt{Qwen3-Coder}} & \textbf{\texttt{gpt-5-mini}} & \textbf{\texttt{llama-4}} \\
\midrule
\multicolumn{5}{c}{\cellcolor{gray!20}\textbf{P2P}} \\

\textsc{Promise} & 0.4044 & 0.3378 & 0.4133 & 0.2889 \\
\textsc{PURE}    & 0.2261 & 0.2522 & 0.2522 & 0.2087 \\
\textsc{SRS}     & 0.2667 & 0.2267 & 0.2133 & 0.2800 \\
\textsc{FQ}      & 0.3600 & 0.2000 & 0.2400 & 0.4200 \\
\midrule
\multicolumn{5}{c}{\cellcolor{gray!20}\textbf{Chebyshev}} \\

\textsc{Promise} & 0.2667 & 0.2578 & 0.2667 & 0.3067 \\
\textsc{PURE}    & 0.5826 & 0.5478 & 0.4783 & 0.5565 \\
\textsc{SRS}     & 0.3733 & 0.2267 & 0.2133 & 0.5600 \\
\textsc{FQ}      & 0.2200 & 0.2400 & 0.2000 & 0.2000 \\
\midrule
\multicolumn{5}{c}{\cellcolor{gray!20}\textbf{RMSE}} \\

\textsc{Promise} & 0.2089 & 0.2489 & 0.2178 & 0.2711 \\
\textsc{PURE}    & 0.2522 & 0.2609 & 0.2609 & 0.2696 \\
\textsc{SRS}     & 0.2400 & 0.2400 & 0.2267 & 0.5867 \\
\textsc{FQ}      & 0.2800 & 0.2200 & 0.2400 & 0.2200 \\
\midrule
\multicolumn{5}{c}{\cellcolor{gray!20}\textbf{IAD}} \\

\textsc{Promise} & 0.2178 & 0.2489 & 0.2178 & 0.2800 \\
\textsc{PURE}    & 0.2522 & 0.2783 & 0.2783 & 0.2696 \\
\textsc{SRS}     & 0.2400 & 0.2400 & 0.2533 & 0.3867 \\
\textsc{FQ}      & 0.2800 & 0.2200 & 0.2200 & 0.2400 \\
\bottomrule
\end{tabularx}
\caption{Cognitive overhead saving achieved by \approach.}
\label{tab:RCO}
\end{table}
To examine the effectiveness of \approach~in reducing cognitive load, we quantify the cognitive load as the number of human interactions required to achieve a goal. To this end, we measure the cognitive overhead saved by \approach~relative to a compared method that permits human interaction. Suppose that we compare \approach~with another method $B$:

\begin{itemize}
    \item For each requirement, we find the smallest interaction count $N_{b,i}$ that $B$ reaches its best metric value $b$ averaged over 5 repeats, then averaging the count over $k$ requirements: $N_{b\text{-avg}} = \frac{1}{k}\sum_{i=1}^k N_{b,i}$.
    \item We then find the smallest interaction count $N_{a,i}$ for \approach~ to reach the same $b$ averaged over 5 repeats on the same requirement, compute the average count over $k$ requirement: $N_{a\text{-avg}} = \frac{1}{k}\sum_{i=1}^k N_{a,i}$.
    \item Quantify the relative cognitive overhead of \approach~ via the ratio $\frac{N_{a\text{-avg}}}{N_{b\text{-avg}}}$.
\end{itemize}

The smaller the ratio, the more savings can be achieved by \approach, e.g., for $k$ requirements, if \texttt{DeepSeek-V3} reaches their best average metric $b$ at average $N_{b-avg}=5$. We find at which average smallest interaction count that \approach~can reach $b$, says $N_{a-avg}=3$; we can finally have ${N_{a-avg}\over N_{b-avg}}= {3 \over 5}=0.6$---\approach~has only 60\% overhead to that of \texttt{DeepSeek-V3}.

The results are in Table~\ref{tab:RCO}: \approach~significantly reduces the cognitive overhead in all cases with as low as $20\%$ of the other method.

\subsection{Sensitivity to the Interaction Rounds $N$}

To understand the sensitivity of \approach~to $N$, we set $N\in[1,2,...,9]$ and plot the mean metric values standardized over all datasets. From Figure~\ref{fig:trade_off}, as expected, a larger $N$ indicates better results, but clearly, more interaction rounds could incur higher cognitive overhead for the stakeholder, which is not ideal. However, we see that the default $N=5$ reaches a well-balanced trade-off between performance and cognitive overhead, which is safer. 

\input{figures/trade_off}

\subsection{Sensitivity to the Changing Step $\Delta$}

\begin{figure}[t!]
    \centering
    \definecolor{color1}{RGB}{114, 188, 213}
    \definecolor{color2}{RGB}{191, 11, 59}
    \definecolor{color3}{RGB}{253, 139, 60}
    \definecolor{color4}{RGB}{255, 217, 118}
    \definecolor{highlightgreen}{RGB}{0, 180, 0}

    \pgfplotsset{
        my_scatter_style/.style={
            width=2.7cm,
            height=2.7cm,
            scaled y ticks=false,
            axis x line = bottom,
            axis y line = left,
              scaled y ticks={base 10:2},
            yticklabel style={/pgf/number format/fixed, /pgf/number format/precision=2},
            xtick={3, 6, 9, 12, 15},
            xmin=4, xmax=16,
            tick label style={font=\tiny},
            label style={font=\scriptsize},
            enlarge y limits={rel=0.2}, 
            every axis plot/.append style={
                only marks, 
                mark=*, 
                mark size=1.5pt, 
                mark options={draw=black, line width=0.7pt}
            }
        }
    }

    \begin{subfigure}[b]{0.22\linewidth}
        \centering
        \begin{tikzpicture}
            \begin{axis}[my_scatter_style, y label style={at={(-0.35,0.5)}}, ylabel={Metric value}, xlabel={$\Delta$ value}, ymin=0.0, ymax=0.6, ytick={0.0, 0.2, 0.4, 0.6}]
                \addplot[color=color1] coordinates {(5,0.335) (6,0.333) (7,0.308) (8,0.286) (9,0.257) (10,0.163) (11,0.255) (12,0.311) (13,0.265) (14,0.308) (15,0.277)};
                \draw[highlightgreen, very thick] (axis cs:10, 0.163) circle (2.5pt);
            \end{axis}
        \end{tikzpicture}
        \caption{P2P}
    \end{subfigure}
    ~\hspace{0.1cm}
    \begin{subfigure}[b]{0.24\linewidth}
        \centering
        \begin{tikzpicture}
            \begin{axis}[my_scatter_style, xlabel={$\Delta$ value}, ymin=-0.1, ymax=0.8, ytick={0.0, 0.4, 0.8}]
                \addplot[color=color2] coordinates {(5,0.452) (6,0.426) (7,0.344) (8,0.371) (9,0.297) (10,0.074) (11,0.273) (12,0.381) (13,0.264) (14,0.362) (15,0.301)};
                \draw[highlightgreen, very thick] (axis cs:10, 0.074) circle (2.5pt);
            \end{axis}
        \end{tikzpicture}
        \caption{Chebyshev}
    \end{subfigure}
    \hfill
     ~\hspace{-0.07cm}
    \begin{subfigure}[b]{0.22\linewidth}
        \centering
        \begin{tikzpicture}
            \begin{axis}[my_scatter_style, xlabel={$\Delta$ value}, ymin=-0.05, ymax=0.5, ytick={0.0, 0.25, 0.5}]
                \addplot[color=color3] coordinates {(5,0.284) (6,0.236) (7,0.204) (8,0.203) (9,0.165) (10,0.030) (11,0.145) (12,0.212) (13,0.133) (14,0.207) (15,0.167)};
                \draw[highlightgreen, very thick] (axis cs:10, 0.030) circle (2.5pt);
            \end{axis}
        \end{tikzpicture}
        \caption{RMSE}
    \end{subfigure}
     ~\hspace{-0.07cm}
    \begin{subfigure}[b]{0.22\linewidth}
        \centering
        \begin{tikzpicture}
            \begin{axis}[my_scatter_style, xlabel={$\Delta$ value}, ymin=-0.05, ymax=0.4, ytick={0.0, 0.2, 0.4}]
                \addplot[color=color4] coordinates {(5,0.238) (6,0.187) (7,0.160) (8,0.151) (9,0.114) (10,0.019) (11,0.113) (12,0.168) (13,0.096) (14,0.160) (15,0.129)};
                \draw[highlightgreen, very thick] (axis cs:10, 0.019) circle (2.5pt);
            \end{axis}
        \end{tikzpicture}
        \caption{IAD}
    \end{subfigure}

    \caption{Sensitivity of \approach~to $\Delta$ over all datasets  (detailed results can be found at Appendix~\ref{sec:full_delta_sen}).}
    \label{fig:delta_trade_off}
\end{figure}
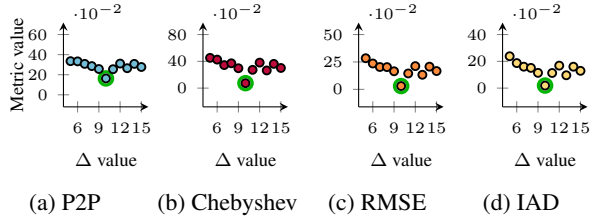

By default, \approach~sets $\Delta = 10\% \times T$ to construct the quantification $f_{t,0}$. To verify the rationality of this heuristic setting, we conduct a sensitivity analysis by performing a parameter sweep for $\Delta$ across all datasets. We vary $\Delta$ from 5\% to 15\% of the threshold $T$ with a step size of 1\%.

As from Figure~\ref{fig:delta_trade_off}, clearly, 10\% is a reasonable setting for $\Delta$, leading to the best results in general. 


\section{Conclusion}

This paper formalizes the problem of quantifying performance requirements and presents a \textit{conjecture and inquiry} approach via interactive retrieval-augmented preference elicitation, dubbed \approach. The key is that \approach~not only performs interactive quantification via retrieval reasoning of preference using problem-specific knowledge, but also does so with minimized stakeholders' cognitive overhead in several aspects. Evaluation against 10 state-of-the-art methods and four real-world datasets reveals that \approach~outperforms the others on both efficiency and efficacy in general.


Looking forward, the above provides a formal problem formulation and solution for quantifying performance requirements, and more importantly, it paves the way towards a research avenue of interactive preference elicitation in the field.


\section*{Limitations}


\textbf{Specific to performance requirement quantification:} \approach~is an approach that is specifically designed for preference reasoning and tuning under software performance requirement quantification, taking many of its characteristics into account. Further, the linear slope in \approach~can be easily replaced by, e.g., a nonlinear sigmoid/error function, and hence the linearity is not a constraint nor a hard assumption. While many of its concepts are general and can be transferred to other problems, this remains subject to future work. 

\textbf{Quantifying initially complex requirements with multiple patterns:} While the three patterns formulated in \approach~are generic enough to represent the practical performance requirements, there exist complex ones that contain multiple patterns. To quantify those, currently \approach~assumes that the initially given requirements need to be fragmented, such that each fragment serves as an independent requirement that naturally contains only one pattern; though they might become more complex following the preference reasoning and interactive preference tuning. Although this limitation implies the need for extra efforts when using \approach, the fragmentation is often straightforward and practically only $22\%$ real-world performance requirements contain more than one pattern/threshold~\cite{DBLP:journals/corr/abs-2511-03421}. hence we anticipate that \approach~would work fine for most of the real-world cases.


\textbf{The availability of past quantification examples:} The retrieval-analogical preference reasoning relies on past examples as the analogy to convert an initial draft of quantification. However, at the beginning, when there are too few accumulated past examples, the analogies to be retrieved are limited. However, this would be relieved as the quantification and reasoning proceed with more performance requirements. Further, it is possible to generate synthetic data, as we did in this work, to serve as the prior knowledge therein.


\textbf{Threats to validity:} Threats to construct validity might arise from the metrics used in the evaluation. To mitigate this, in this work, we use all the common metrics that measure the distance between the produced quantification and the ground truth. Also, for all experiments, the setting of parameters might affect internal validity. We have considered that by using the same/default settings for state-of-the-art methods, while examining the hyperparameter $N$ for \approach; we have also maintained consistency on all other setups for better fairness. Yet, admittedly, we cannot ensure that those are optimal for all cases. Finally, the subject datasets and methods compared could cause threats to external validity. Although we have examined 10 state-of-the-art methods, covering four diverse categories, and under four widely-known available real-world performance requirements datasets (as there are not much publicly available datasets for the problem), it remains difficult to guarantee that the same results would always be observed in all new cases. As such, we acknowledge that examining more datasets/methods, if publicly available, might prove more fruitful.

\section*{Ethical Considerations}

All human-involved processes in this paper, i.e., the manual annotation of the most preferred quantification of the requirements in the datasets and the interactive adjustments in experiments, are strictly conducted in compliance with the ethical standards for academic research. The participants in this study are authors and collaborators of this paper/work, who are experienced software engineers, and their involvement is carried out in full accordance with legal and regulatory policies, limiting to content related to the performance requirements only. Notably, the participants’ legitimate right---includes the right to withdraw from the study at any time---has been fully protected. 


The real-world performance requirement dataset used in this study, as well as the synthetic data generated via \texttt{GPT-4}, do not involve any personal privacy information, commercially sensitive data, or confidential information targeting specific users. The requirement content mainly focuses on general performance metrics in software engineering (e.g., response time, throughput, and resource utilization). Furthermore, all open-source LLMs employed in this study (e.g., \texttt{Qwen} and {GPT-2}) are used in strict compliance with their respective open-source licenses and usage policies. 

We hereby commit that all steps of data collection/labeling, data generation, and human-machine interaction in this study are fully aligned with scientific research ethics.

\section*{Acknowledgment}

This work was supported by a NSFC Grant (62372084).


\bibliography{custom}


\appendix

\section*{Appendix}

\section{Additional Details for Retrieval-Generative Quantification}
\label{append:sft_samples}

\begin{table*}[t!]
    \centering
 
    \renewcommand{\arraystretch}{1.3}
    
    
    \begin{tabularx}{\linewidth}{l >{\raggedright\arraybackslash}X} 
        \toprule
        \textbf{Pattern Type} & \textbf{Anchor Phrase} \\ 
        \midrule
        \textbf{$P_1$}  & \textit{``no less than''}, \textit{``at least''}, \textit{``greater than''}, \textit{``minimum of''}, \textit{``not below''}, \textit{``above''}, \textit{``exceeding''}, \textit{``no fewer than''}, \textit{``greater than or equal to''}, \textit{``at minimum''} \\ 
        \addlinespace
        \textbf{$P_2$}  & \textit{``no more than''}, \textit{``at most''}, \textit{``less than''}, \textit{``maximum of''}, \textit{``not exceeding''}, \textit{``under''}, \textit{``below''}, \textit{``up to''}, \textit{``at maximum''}, \textit{``with in''} \\ 
        \addlinespace
        \textbf{$P_3$}  & \textit{``exactly''}, \textit{``equal to''}, \textit{``precisely''}, \textit{``specifically''}, \textit{``fixed at''}, \textit{``set to''}, \textit{``equivalent to''}, \textit{``identical to''}, \textit{``precisely at''}, \textit{``designated as''} \\
        \bottomrule
    \end{tabularx}
       \caption{The anchor phrases for all pattern types.}
    \label{tab:anchor}
\end{table*}

\subsection{Full Anchor List in Retrieval-based Classification}
\label{append:anchor_phrases}

Here, the anchor phrases serve as important domain-specific knowledge for incorporating label semantics in the retrieval-based classification. To that end, we extract and prepare the anchors following the steps below:

\begin{enumerate}
    \item Select highly representative and relevant phrases from the keyword sets of performance requirements summarized in prior work~\cite{DBLP:journals/corr/abs-2511-03421}.
    \item Drawing on those, supplement more synonymous phrases via the English lexical database.
    \item Carefully assign those phrases into the correct pattern types through agreements among all authors. To balance the effort and representativeness, we cap 10 phrases for each pattern.
\end{enumerate}

Table~\ref{tab:anchor} lists all the 30 anchor phrases corresponding to the three defined patterns. Note that when fine-tuning the RoBERTa, all requirements and anchors are used as the training samples.



\subsection{Sample/Prompt in Generative Threshold Extraction}
\label{sec:thres}

To extract the relevant threshold $T$ in a requirement, \approach~fine-tunes the lightweight GPT-2 with only 774M parameters. Each sample consists of an instruction-based prompt and the expected ground truth of the threshold, e.g.:

\begin{halfpromptbox}{Fine tuning sample}
\textbf{Input:} Please extract the numeric threshold from the following performance requirements: ``\textit{The response time must not exceed 200ms.}''\\
\textbf{Output:} 200
\end{halfpromptbox}

Upon extraction, we use the same prompt to obtain the threshold of the target requirement, e.g.:

\begin{halfpromptbox}{Prompt}
Please extract the numeric threshold from the following performance requirements.\\ 
\textbf{Requirement:} ``\textit{The system should support at least 500 concurrent stakeholders for 7 days, 24 hours.}''
\end{halfpromptbox}


\begin{algorithm}[t!]
\caption{\textsc{Retrieval\_Analogical\_ Preference\_Reasoning}}
\label{alg:diff_reasoning}
\footnotesize
\begin{algorithmic}[1]
\State \textbf{Input:} Performance requirement $s_t$, initial draft quantification for the target $f_{t,0}$ and set of past quantification examples $\mathcal{S}:\{s_i = \{f_{i,0},f_{i}^*\}| i \in [1,t-1]\}$

\State \textbf{Output:} Converted/reasoned quantification $f'_{t,0}$


\State $s_k=\{f_{k,0},f_{k}^*\} \leftarrow \arg\max_{s_i \in \mathcal{S}, |f_{i,0}| = |f_{t,0}|}$ $\textsc{semantic\_sim}(s_t, s_i)$

\State $\mathcal{O} = \emptyset$

\Statex \(\triangleright\) Points alignment
\State $\mathcal{M} \leftarrow$ \textsc{matching\_By\_KM}($f_{k,0}$,$f_{k}^*$)

\If{points in $f_{k,0}$ > points in $f_{k}^*$}
       \State $\mathcal{O} \leftarrow \mathcal{O} \cup \{\textsc{remove}(u) \mid u \in f_{k,0}; u \notin \mathcal{M}\}$
\ElsIf{points in $f_{k,0}$ < points in $f_{k}^*$}
\State $\mathcal{O} \leftarrow \mathcal{O} \cup \{\textsc{add}(u) \mid u \in f_{k}^*; u \notin \mathcal{M}\}$ 
\EndIf

\Statex \(\triangleright\) Changes identification
\State $f'_{k,0} \leftarrow \text{apply } \mathcal{O} \text{ to } f_{k,0}$
\State $\mathcal{M}' \leftarrow$ \textsc{matching\_By\_KM}($f'_{k,0}$,$f_{k}^*$)
\For{each matched pair $(u_i, v_j) \in \mathcal{M}'$}
    \If{$u_i \neq v_j$}
        \State $\mathcal{O} \leftarrow \mathcal{O} \cup \{\textsc{change}(u_i, v_j)\}$
    \EndIf
\EndFor

\Statex \(\triangleright\) Operations sequencing
\State Place the \textsc{change} in $\mathcal{O}$ as after \textsc{add} and \textsc{remove}

\State $f'_{t,0} \leftarrow \text{apply } \mathcal{O} \text{ to } f_{t,0}$
\State \textbf{return} $f'_{t,0}$
\end{algorithmic}
\end{algorithm}

\section{Additional Details for Retrieval-Analogical Preference Reasoning}
\label{append:analogical_reasoning}
The pseudo code for the retrieval-analogical preference reasoning has been shown in Algorithm~\ref{alg:diff_reasoning}, where the key steps discussed are highlighted. The optimal matching via the KM algorithm has been illustrated as the pseudo code in Algorithm~\ref{alg:matching}.

\begin{algorithm}[t!]
\caption{\textsc{matching\_By\_KM}}
\label{alg:matching}
\footnotesize
\begin{algorithmic}[1]
\State \textbf{Input:} Initial points $f_{k,0} : \mathcal{U}=\{u_1, \dots, u_n\}$; final points $f_{k}^* : \mathcal{V}=\{v_1, \dots, v_m\}$
\State \textbf{Output:} Optimal matching $\mathcal{M}$
\State $\mathcal{W} \leftarrow$ a $n \times m$ matrix
\For{$i \leftarrow 1$ \textbf{to} $n$}
\For{$j \leftarrow 1$ \textbf{to} $m$}
\State $\mathcal{W} \leftarrow w_{ij} = - \sqrt{(x_{u_i} - x_{v_j})^2 + (y_{u_i} - y_{v_j})^2}$ 
\EndFor
\EndFor
\State $\mathcal{M} \leftarrow \text{KMAlgorithm}(\mathcal{W})$ \Comment{Solving max weight matching}
\State \textbf{return} $\mathcal{M}$
\end{algorithmic}
\end{algorithm}

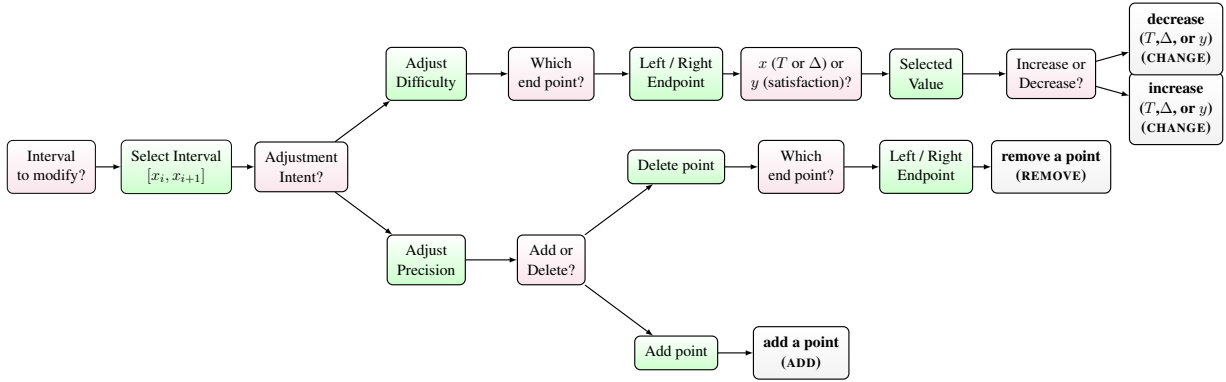
\begin{figure*}[htbp]
    \centering
    \begin{adjustbox}{width=\linewidth}
    \begin{tikzpicture}
      [
        grow                    = right,
        sibling distance        = 12em, 
        level distance          = 8.0em,
        edge from parent/.style = {draw, -latex, thick}, 
        sloped,
        treenode/.style = {shape=rectangle, rounded corners, draw, align=center, font=\normalsize, inner sep=7pt},
        question/.style = {treenode, top color=white, bottom color=purple!10},
        option/.style   = {treenode, top color=white, bottom color=green!20},
        atomic/.style   = {treenode, font=\normalsize\bfseries, top color=white, bottom color=gray!10},
        dots/.style     = {draw=none, rectangle, align=center, font=\Large, inner sep=2pt}
      ]
    
      \node [question] {Interval\\to modify?}
        child { node [option] {Select Interval\\$[x_i, x_{i+1}]$}
          child { node [question] {Adjustment\\Intent?}
            child { node [option] {Adjust\\Precision} 
              child { node [question] {Add or\\Delete?}
                child { node [option] {Add point}
                  child { node [atomic] {add a point\\(\textsc{add})} }
                }
                child { node [option] {Delete point}
                  child { node [question] {Which\\end point?}
                    child {
                        node [option] {Left / Right\\Endpoint}
                        child {
                            node [atomic] {remove a point\\(\textsc{remove})} 
                        }
                    }
                  }
                }
              }
            }
            child { node [option] {Adjust\\Difficulty} 
              child { node [question] {Which\\end point?}
                child { node [option] {Left / Right\\Endpoint}
                  child { node [question] {$x$ ($T$ or $\Delta$) or\\$y$ (satisfaction)?}
                    child { node [option] {Selected\\Value}
                      child{ node [question] {Increase or\\Decrease?}
                      [sibling distance=4.5em]
                        child { node [atomic] {increase \\ ($T$,$\Delta$, or $y$)\\(\textsc{change})} } 
                         child { node [atomic] {decrease \\ ($T$,$\Delta$, or $y$)\\(\textsc{change})} } 
                      }
                    }
                  }
                }
              }
            }
          }
        };
    \end{tikzpicture}
    \end{adjustbox}
    \caption{The complete question tree in \approach~for interaction.}
    \label{fig:questionnaire}
\end{figure*}

\section{Full Question Tree of Interactive Preference Tuning}
\label{append:interactive_system}

The complete tree-based multiple-choice questions used in interactive preference tuning for question-answering are shown in Figure \ref{fig:questionnaire}. In particular, the designs follow three levels of hierarchy to progressively prompt the stakeholder for intuitive and iterative feedback.

Note that, in \approach, completing one round means answering all questions from the root to one of the leaves, leading to one operation; the number of rounds is bounded by the hyperparameter $N=5$. In all experiments, we have rarely observed that the quantification matches the ground truth with fewer than 5 interaction rounds.

\section{Other Experiment Setting Details}

\begin{table*}[t!] 

    \begin{promptbox}{Prompt used for vanilla LLMs}
    For a performance requirement statement, we can quantify it. Specifically, we expect to derive a function expression that takes a performance metric as input and outputs the stakeholder satisfaction level corresponding to that metric. We may simply assume that the function expression is a piecewise linear function. For example, given the requirement statement:
    
    > In the scenario of real-time electrocardiogram (ECG) monitoring, the software shall receive and process ECG signal data at a sampling frequency no lower than 1000 Hz.
    
    The corresponding piecewise linear quantification function is:
    $$
    f(x) = 
    \begin{cases}
    0 & \text{if } x \leq 900 \\
    1/100(x-900) & \text{if } 900 < x < 1000 \\
    1 & \text{if } x \geq 1000
    \end{cases}
    $$
    A piecewise linear function has multiple "inflection points". The set of inflection points for the above piecewise linear function is:
    $$
    \{(900,0),(1000,1)\}
    $$
    
    However, this form of quantification is imprecise. You need to ask questions to the requirement setter to adjust the quantification form. The quantification form shall be uniformly represented in the form of a point list. You may ask the requirement setter about the aspects of the current quantification form that they are dissatisfied with, and then adjust the quantification form based on their responses. Your inquiries are subject to constraints: you are not allowed to directly ask for the exact ideal quantification form they have in mind. Instead, you can only ask ambiguous questions, such as whether the coordinate value of a certain point is too large or too small (you cannot directly ask for the specific coordinate value), and whether the number of segments is too many or too few (you cannot directly ask for the exact number of segments). Within these constraints, you may ask any questions you want. Your goal is to guess the quantification form that satisfies the stakeholder. Note that you are only allowed to ask 5 questions in total, with one question per round. After I have answered your 5th question, you need to output the final confirmed quantification form. The performance requirement quantification task you need to handle is as follows:
    \end{promptbox}
    \caption{Prompt used for vanilla LLMs.}
    \label{tab:prompt_bot}
\end{table*}

\subsection{Additional Details of Evaluation Metrics}
\label{sec:app_m}

The specific formulas for calculating the evaluation metrics are discussed below:

\begin{itemize}
    \item \textbf{Point-to-Point Distance (P2P)}: This metric assesses the structural and positional similarity of the functions' inflection point sets. We normalize the domain to $[0, 1]$. Optimal matching is achieved via the KM algorithm~\cite{DBLP:books/daglib/p/Kuhn10}, accumulating the Euclidean distances of matched pairs:
    $$
    \small
    \text{P2P} =  \sum_{(u_i, v_j) \in \mathcal{M}} w_{ij}
    $$

    \item \textbf{Maximum Deviation (Chebyshev)}: This metric quantifies the worst-case vertical discrepancy between the two quantification. It is calculated as the maximum absolute difference between function values across the domain, determined by searching all inflection points, endpoints, and sampled intervals:
    $$
    \small
    \text{Chebyshev} = \max_x |f_1(x) - f_{2}(x)|
    $$

    \item \textbf{Root Mean Square Error (RMSE) for $y$}: This metric measures the robust average functional deviation along the curve. After normalizing the domain to $[0, 1]$, we select $N$ uniformly sampled points $x_i$ and compute the root mean square of the function value differences:
    $$
    \small
    \text{RMSE} = \sqrt{\frac{1}{N}\sum_{i=1}^{N}(f_{1}(x_i) - f_{2}(x_i))^2}
    $$

    \item \textbf{Integrated Area Difference (IAD)}: This metric evaluates the overall difference in the implied requirement difficulty, as the integral area represents this dimension. It is the absolute difference between the areas enclosed by two functions and the x-axis, within the normalized common domain $[0, 1]$:
    $$
    \small
    \text{IAD} = \left| \int f_1(x) \, dx - \int f_2(x) \, dx \right|
    $$
\end{itemize}

\subsection{Hyperparameters}

For all state-of-the-art methods, their hyperparameters are set as the default; for \approach, the hyperparameters are also used the default, e.g., $\tau=0.07$ for fine-tuning the GPT-2. The number of interactions $N$ is set to 5 unless otherwise stated.

\subsection{Prompt used for vanilla LLMs}
\label{sec:llm}
The prompt used for vanilla LLMs is shown in Table \ref{tab:prompt_bot}.

\subsection{Prompt used for RAG-based methods}
\label{sec:rag}
The prompt used for RAG-based methods is shown in Table \ref{tab:prompt_rag}. 


\begin{table*}[t!] 

    \begin{promptbox}{Prompt used for RAG-based methods}
    Accurately converting performance requirements described in natural language into computable quantitative indicators is a key challenge in the field of software engineering, shifting from qualitative evaluation to quantitative analysis. Our goal is to establish a stakeholder satisfaction function $f(x)$ for any performance indicator $x$, where the function outputs the stakeholder satisfaction (ranging from $[0, 1]$) corresponding to the value of $x$.\\
    
    We adopt piecewise linear functions to characterize such satisfaction curves, as they offer simplicity, ease of interpretation, and sufficient expressive power. This type of piecewise linear function can be uniquely determined by its set of inflection points $\{(x_1, y_1), \ldots, (x_n, y_n)\}$.\\
    
    For example, consider the requirement statement:
    
    > In real-time ECG monitoring scenarios, the software must receive and process ECG signal data at a sampling frequency of no less than 1000Hz.
    
    With a preset tolerance range of $10\%$, the set of inflection points for this requirement is:
    $$
    \{(900, 0), (1000, 1)\}
    $$
    This set can directly restore the corresponding quantitative function $f(x)$:
    $$
    f(x) = 
    \begin{cases}
    0 & \text{if } x \leq 900 \\
    1/100(x-900) & \text{if } 900 < x < 1000 \\
    1 & \text{if } x \geq 1000
    \end{cases}
    $$
    Thus, our task is formally defined as a sequence-to-sequence conversion problem:
    $$
    \text{Performance Requirement Statement} \rightarrow [(x_1, y_1), \ldots, (x_n, y_n)]
    $$
    
    However, the quantitative form directly derived from performance requirement statements is not precise and does not fully align with the quantitative form expected by the requirement setters. The following examples illustrate the differences between the quantitatively derived form from the literal meaning and the expected quantitative form by the requirement setters:
    $$
    [\text{Place the retrieved historical samples here}]
    $$
    Please refer to the above difference samples. **Note that you need to analyze the differences from the base form to the prefer form and apply them to your current task** to provide a more preference-aligned quantitative form (prefer form) for the following performance requirement. The base form is provided; you need to infer the prefer form. (Output the result only in the form of a list of points, and submit just the final prefer form):
    \end{promptbox}
    \caption{Prompt for RAG-based methods.}
    \label{tab:prompt_rag}
\end{table*}

\subsection{Prompt used for preference-optimized methods with RL}
\label{sec:rl}

The prompt used for RAG-based methods is shown in Table \ref{tab:prompt_RLHF}. Note that the prompts are optimized via reinforcement learning to make the quantitative forms generated by LLMs more aligned with the preferences of stakeholders.

\begin{table*}[t!] 

    \begin{promptbox}{Prompt used for preference-optimized methods with RL}
    Accurately converting performance requirements described in natural language into computable quantitative indicators is a key challenge in the field of software engineering, transitioning from qualitative evaluation to quantitative analysis. Our goal is to establish a stakeholder satisfaction function \( f(x) \) for any performance indicator \( x \), where the function outputs the stakeholder satisfaction (within the range \([0, 1]\)) corresponding to the value of \( x \).\\
    
    We adopt a piecewise linear function to characterize this satisfaction curve, which offers simplicity, interpretability, and sufficient expressive power. Such a piecewise linear function can be uniquely determined by its set of inflection points \(\{(x_1, y_1), \ldots, (x_n, y_n)\}\).\\
    
    For example, consider the requirement statement:
    
    > In real-time ECG monitoring scenarios, the software must receive and process ECG signal data at a sampling frequency of no less than 1000Hz.
    
    Given a preset tolerance range of 10
    $$
    \{(900, 0), (1000, 1)\}
    $$
    This set can directly restore the corresponding quantitative function \( f(x) \):
    $$
    f(x) = 
    \begin{cases}
    0 & \text{if } x \leq 900 \\
    1/100(x-900) & \text{if } 900 < x < 1000 \\
    1 & \text{if } x \geq 1000
    \end{cases}
    $$
    Thus, our task is formally defined as a sequence-to-sequence conversion problem:
    $$
    \text{Performance requirement statement} \rightarrow [(x_1, y_1), \ldots, (x_n, y_n)]
    $$
    
    Please convert the following performance requirement into a quantitative form based on the above theory (output the result as a list of 2D points, such as [[10.0, 1.0], [11.0, 0.0]]. Please strictly follow the specified format for output, and do not include any additional content.):
    \end{promptbox}
    \caption{Prompt for preference-optimized methods with RL.}
    \label{tab:prompt_RLHF}
\end{table*}

\section{Additional Qualitative Case Study}
\label{sec:add_quanlitative}

\noindent \textbf{Commonly effective cases for \approach:} A typical kind of performance requirements that \approach~handles well is: \textit{``The system requests per second (req/s) shall support at least 200.''} 
The stakeholder's most preferred quantification is $f^*_{t}: \{(180, 0), (195, 0.6), (198, 0.8), (200, 1)\}$. 
\approach~initially generates $f_{t,0}: \{(180, 0), (200, 1)\}$. 
Subsequently, the retrieval-analogical preference reasoning phase retrieves a relevant historical example $s_k$: \textit{``The number of concurrent users shall reach 100,''} 
which followed the trajectory $f_{k,0}: \{(90, 0), (100, 1)\} \to f_k^*: \{(98, 0), (100, 1)\}$, as the most appropriate analogy, i.e., stricter quantification (higher difficulty) is more likely to be preferred by the stakeholder. 
As such, \approach~then accordingly convert $f_{t,0}$ into the reasoned state $f'_{t,0}: \{(195, 0), (200, 1)\}$. 
In the interaction, \approach~guides the stakeholder through the question tree illustrated in Figure~\ref{fig:questionnaire}:
\begin{itemize}
    \item \textbf{Round 1 (Adjusting Difficulty):} The stakeholder finds the reasoning result too strict and selects the interval $[196, 200]$. 
          The path is: \textsc{Left Endpoint} $\to$ \textsc{Select x} $\to$ \textsc{Decrease}. 
          Applying the step size of $10\%$, the $x$ value decreases from $196$ to $175.5$. 
          Thus, the current state is: $\{(175.5, 0), (200, 1)\}$.
    \item \textbf{Round 2 (Adjusting Difficulty):} Finding $175.5$ slightly too loose/relaxed, the stakeholder reverses the direction to \textsc{Increase} $x$. 
          Due to the direction reversal on the same point, the step size attenuates to $5\%$. 
          The $x$ value increases to $175.5 \times (1 + 5\%) = 184.275$, leading to the current state: $\{(184, 0), (200, 1)\}$.
    \item \textbf{Round 3 (Adjusting Precision):} The stakeholder adds a transitional satisfaction level to increase the precision of the curve. 
          The path is: \textsc{Interval to modify} $\to$ \textsc{Adjust Precision} $\to$ \textsc{Add point}. 
          \approach~inserts a new point at the mean of $[184, 200]$, resulting in $(192, 0.5)$. 
          Thus, the current state is: $\{(184, 0), (192, 0.5), (200, 1)\}$.
    \item \textbf{Round 4 (Adjusting Precision):} The stakeholder decides to further increase the precision by adding a point between the intermediate point and the endpoint. 
          The path is: \textsc{Interval to modify} $\to$ \textsc{Adjust Precision} $\to$ \textsc{Add point}. 
          \approach~inserts a new point at the mean of $[192.138 , 200]$, resulting in $(196, 0.75)$.
          The current state becomes: $\{(184, 0), (192, 0.5), (196, 0.75), (200, 1)\}$.
    \item \textbf{Round 5 (Adjusting Difficulty):} The stakeholder further refines the satisfaction level $y$ for the intermediate point. 
          The path is: \textsc{Adjust Difficulty} $\to$ \textsc{Select y} $\to$ \textsc{Increase}. 
          The $y$ value increases by $10\%$ to $0.5 \times (1 + 10\%) = 0.55$. 
          Thus, the current state is: $\{(18, 0),(192, 0.55), (196, 0.75), (200, 1)\}$.
\end{itemize}

Finally, \approach~reaches a result relatively close to $f^*_{t}$ through five rounds of tree path question-answering based on the question tree.

In contrast, the domain-specific method (\texttt{LQPR}) enables automated mapping from natural language to quantitative forms; however, it essentially relies on static rule matching and completely neglects the dynamic impacts of stakeholders' subjective preferences on the tolerance margin $\Delta$ and satisfaction curves: it at most output a quantification similar to $f_{t,0}: \{(180, 0), (200, 1)\}$.


Although the vanilla LLMs support interactive capabilities, without the guidance of the quantitative theoretical framework proposed in this paper, the active inquiries generated by LLMs tend to be divergent and fail to capture the key aspects of stakeholder preferences. For instance, the result obtained using \texttt{Qwen3-coder} is: $\{(199, 0), (200, 1)\}$, which oversimplifies the stakeholder's complex satisfaction decay into a binary threshold and fails to explore the tolerance margin between 180 and 200 req/s.

RAG-based methods incorporate historical contextual information, yet they overly depend on the model's inherent reasoning capabilities to interpret stakeholder preferences. Such an analysis approach is both unstable and indirect. In this case, it produces inconsistent and unstable numerical mappings $\{(180, 0), (192, 0.8), (200, 1)\}$, where the satisfaction values are biased by loosely related historical context rather than the actual fine-grained preferences elicited through structured interaction.

\begin{table*}[t!]
\centering
\scriptsize

\begin{tabular}{l c c c c | c c c c}
\toprule
\textbf{$N$ value} & P2P & Chebyshev & RMSE & IAD & P2P & Chebyshev & RMSE & IAD \\
\midrule
& \multicolumn{4}{c}{\cellcolor{gray!20}\textbf{\textit{\textsc{Promise} Dataset}}} & \multicolumn{4}{c}{\cellcolor{gray!20}\textbf{\textit{\textsc{PURE} Dataset}}} \\
\cmidrule(lr){2-5} \cmidrule(lr){6-9} 
1 & 0.315 (0.145) & 0.038 (0.011) & 0.018 (0.002) & 0.012 (0.001) & 0.108 (0.010) & 0.257 (0.069) & 0.133 (0.017) & 0.104 (0.011) \\
2 & 0.284 (0.147) & 0.034 (0.008) & 0.017 (0.002) & 0.013 (0.001) & 0.073 (0.003) & 0.180 (0.041) & 0.096 (0.006) & 0.073 (0.002) \\
3 & 0.264 (0.147) & 0.028 (0.008) & 0.016 (0.002) & 0.010 (0.001) & 0.059 (0.005) & 0.184 (0.045) & 0.097 (0.009) & 0.066 (0.005) \\
4 & 0.252 (0.148) & 0.028 (0.013) & 0.016 (0.003) & 0.011 (0.001) & 0.044 (0.004) & 0.117 (0.039) & 0.060 (0.006) & 0.052 (0.003) \\
\textbf{5} & \textbf{0.238 (0.384)} & \textbf{0.032 (0.111)} & \textbf{0.014 (0.048)} & \textbf{0.011 (0.038)} & \textbf{0.040 (0.052)} & \textbf{0.131 (0.210)} & \textbf{0.053 (0.082)} & \textbf{0.033 (0.048)} \\
6 & 0.244 (0.149) & 0.022 (0.006) & 0.010 (0.001) & 0.010 (0.001) & 0.023 (0.003) & 0.126 (0.048) & 0.065 (0.008) & 0.041 (0.003) \\
7 & 0.243 (0.148) & 0.031 (0.012) & 0.016 (0.002) & 0.011 (0.001) & 0.025 (0.003) & 0.104 (0.037) & 0.047 (0.006) & 0.032 (0.002) \\
8 & 0.241 (0.149) & 0.024 (0.013) & 0.015 (0.003) & 0.010 (0.001) & 0.025 (0.002) & 0.117 (0.054) & 0.043 (0.008) & 0.028 (0.003) \\
9 & 0.237 (0.150) & 0.018 (0.005) & 0.009 (0.001) & 0.007 (0.001) & 0.022 (0.003) & 0.097 (0.039) & 0.042 (0.006) & 0.026 (0.002) \\
\midrule
\midrule
& \multicolumn{4}{c}{\cellcolor{gray!20}\textbf{\textit{\textsc{SRS} Dataset}}} & \multicolumn{4}{c}{\cellcolor{gray!20}\textbf{\textit{\textsc{FQ} Dataset}}} \\
\cmidrule(lr){2-5} \cmidrule(lr){6-9} 
1 & 0.138 (0.048) & 0.143 (0.052) & 0.058 (0.008) & 0.036 (0.003) & 0.270 (0.170) & 0.000 (0.000) & 0.000 (0.000) & 0.000 (0.000) \\
2 & 0.121 (0.049) & 0.156 (0.055) & 0.063 (0.008) & 0.036 (0.003) & 0.270 (0.170) & 0.000 (0.000) & 0.000 (0.000) & 0.000 (0.000) \\
3 & 0.105 (0.049) & 0.111 (0.029) & 0.044 (0.005) & 0.026 (0.002) & 0.270 (0.170) & 0.000 (0.000) & 0.000 (0.000) & 0.000 (0.000) \\
4 & 0.101 (0.049) & 0.146 (0.035) & 0.064 (0.005) & 0.027 (0.002) & 0.270 (0.170) & 0.000 (0.000) & 0.000 (0.000) & 0.000 (0.000) \\
\textbf{5} & \textbf{0.094 (0.221)} & \textbf{0.087 (0.174)} & \textbf{0.034 (0.069)} & \textbf{0.021 (0.042)} & \textbf{0.270 (0.412)} & \textbf{0.000 (0.000)} & \textbf{0.000 (0.000)} & \textbf{0.000 (0.000)} \\
6 & 0.109 (0.049) & 0.123 (0.047) & 0.054 (0.007) & 0.028 (0.002) & 0.270 (0.170) & 0.000 (0.000) & 0.000 (0.000) & 0.000 (0.000) \\
7 & 0.104 (0.049) & 0.134 (0.050) & 0.065 (0.008) & 0.034 (0.003) & 0.270 (0.170) & 0.000 (0.000) & 0.000 (0.000) & 0.000 (0.000) \\
8 & 0.101 (0.049) & 0.121 (0.062) & 0.063 (0.010) & 0.036 (0.003) & 0.270 (0.170) & 0.000 (0.000) & 0.000 (0.000) & 0.000 (0.000) \\
9 & 0.099 (0.049) & 0.136 (0.063) & 0.063 (0.010) & 0.036 (0.004) & 0.270 (0.170) & 0.000 (0.000) & 0.000 (0.000) & 0.000 (0.000) \\
\bottomrule
\end{tabular}
\caption{Detailed sensitivity analysis of \approach~with varying $N$ value.}
\label{tab:n_sensitivity_res}
\end{table*}
\begin{table*}[t!]
\centering
\scriptsize

\begin{tabular}{l c c c c | c c c c}
\toprule
\textbf{$\Delta$ value} & P2P & Chebyshev & RMSE & IAD & P2P & Chebyshev & RMSE & IAD \\
\midrule
& \multicolumn{4}{c}{\cellcolor{gray!20}\textbf{\textit{\textsc{Promise} Dataset}}} & \multicolumn{4}{c}{\cellcolor{gray!20}\textbf{\textit{\textsc{PURE} Dataset}}} \\
\cmidrule(lr){2-5} \cmidrule(lr){6-9} 
5\%  & 0.419 (0.515) & 0.436 (0.358) & 0.296 (0.309) & 0.257 (0.303) & 0.153 (0.220) & 0.338 (0.268) & 0.168 (0.134) & 0.111 (0.107) \\
6\%  & 0.422 (0.472) & 0.441 (0.310) & 0.273 (0.243) & 0.232 (0.244) & 0.120 (0.076) & 0.280 (0.197) & 0.137 (0.090) & 0.086 (0.075) \\
7\%  & 0.454 (0.501) & 0.471 (0.292) & 0.303 (0.274) & 0.260 (0.280) & 0.157 (0.215) & 0.265 (0.200) & 0.137 (0.113) & 0.100 (0.105) \\
8\%  & 0.422 (0.477) & 0.426 (0.283) & 0.271 (0.245) & 0.225 (0.246) & 0.122 (0.060) & 0.318 (0.184) & 0.155 (0.083) & 0.101 (0.072) \\
9\%  & 0.404 (0.463) & 0.365 (0.277) & 0.232 (0.222) & 0.183 (0.212) & 0.097 (0.047) & 0.272 (0.211) & 0.123 (0.085) & 0.071 (0.052) \\
\textbf{10\%} & \textbf{0.239 (0.385)} & \textbf{0.029 (0.105)} & \textbf{0.014 (0.052)} & \textbf{0.010 (0.042)} & \textbf{0.041 (0.054)} & \textbf{0.136 (0.218)} & \textbf{0.055 (0.085)} & \textbf{0.034 (0.050)} \\
11\% & 0.355 (0.437) & 0.256 (0.277) & 0.151 (0.215) & 0.121 (0.209) & 0.096 (0.077) & 0.281 (0.206) & 0.135 (0.105) & 0.100 (0.092) \\
12\% & 0.388 (0.459) & 0.369 (0.315) & 0.210 (0.225) & 0.175 (0.214) & 0.108 (0.073) & 0.283 (0.212) & 0.121 (0.087) & 0.078 (0.072) \\
13\% & 0.354 (0.402) & 0.254 (0.256) & 0.126 (0.148) & 0.089 (0.137) & 0.097 (0.075) & 0.287 (0.234) & 0.130 (0.096) & 0.076 (0.049) \\
14\% & 0.377 (0.428) & 0.315 (0.248) & 0.178 (0.168) & 0.132 (0.160) & 0.135 (0.078) & 0.332 (0.197) & 0.158 (0.085) & 0.106 (0.069) \\
15\% & 0.381 (0.420) & 0.306 (0.267) & 0.174 (0.201) & 0.137 (0.197) & 0.154 (0.212) & 0.333 (0.243) & 0.167 (0.130) & 0.122 (0.103) \\
\midrule
\midrule
& \multicolumn{4}{c}{\cellcolor{gray!20}\textbf{\textit{\textsc{SRS} Dataset}}} & \multicolumn{4}{c}{\cellcolor{gray!20}\textbf{\textit{\textsc{FQ} Dataset}}} \\
\cmidrule(lr){2-5} \cmidrule(lr){6-9} 
5\%  & 0.223 (0.316) & 0.435 (0.265) & 0.243 (0.207) & 0.193 (0.210) & 0.545 (0.505) & 0.597 (0.354) & 0.428 (0.318) & 0.392 (0.315) \\
6\%  & 0.198 (0.252) & 0.334 (0.252) & 0.173 (0.134) & 0.125 (0.118) & 0.592 (0.437) & 0.650 (0.306) & 0.361 (0.186) & 0.305 (0.171) \\
7\%  & 0.237 (0.331) & 0.345 (0.241) & 0.167 (0.122) & 0.105 (0.104) & 0.385 (0.517) & 0.294 (0.312) & 0.210 (0.274) & 0.175 (0.266) \\
8\%  & 0.199 (0.259) & 0.344 (0.218) & 0.172 (0.116) & 0.112 (0.104) & 0.401 (0.447) & 0.397 (0.261) & 0.212 (0.142) & 0.165 (0.125) \\
9\%  & 0.171 (0.234) & 0.288 (0.194) & 0.139 (0.091) & 0.083 (0.075) & 0.355 (0.479) & 0.261 (0.278) & 0.164 (0.201) & 0.119 (0.183) \\
\textbf{10\%} & \textbf{0.100 (0.221)} & \textbf{0.130 (0.238)} & \textbf{0.051 (0.094)} & \textbf{0.030 (0.056)} & \textbf{0.270 (0.412)} & \textbf{0.000 (0.000)} & \textbf{0.000 (0.000)} & \textbf{0.000 (0.000)} \\
11\% & 0.160 (0.211) & 0.259 (0.190) & 0.115 (0.094) & 0.081 (0.083) & 0.409 (0.392) & 0.294 (0.281) & 0.178 (0.183) & 0.152 (0.169) \\
12\% & 0.171 (0.259) & 0.340 (0.219) & 0.166 (0.118) & 0.104 (0.109) & 0.577 (0.467) & 0.533 (0.290) & 0.349 (0.196) & 0.315 (0.184) \\
13\% & 0.151 (0.209) & 0.278 (0.245) & 0.128 (0.098) & 0.092 (0.069) & 0.456 (0.427) & 0.238 (0.266) & 0.147 (0.200) & 0.128 (0.184) \\
14\% & 0.240 (0.349) & 0.332 (0.192) & 0.190 (0.135) & 0.142 (0.134) & 0.480 (0.442) & 0.470 (0.260) & 0.302 (0.189) & 0.259 (0.190) \\
15\% & 0.182 (0.203) & 0.290 (0.180) & 0.153 (0.091) & 0.110 (0.076) & 0.389 (0.443) & 0.273 (0.194) & 0.174 (0.163) & 0.148 (0.156) \\
\bottomrule
\end{tabular}
\caption{Detailed sensitivity analysis of \approach~with varying $\Delta$ values.}
\label{tab:delta_sensitivity}
\end{table*}

Preference-optimized methods with RL struggle to accurately capture fine-grained numeric preferences when faced with sparse interactive data, and even learn incorrect preferences due to overfitting noisy data, ultimately resulting in counterproductive quantitative outcomes. For this example, \texttt{DPO} generates a non-monotonic quantification $ \{(180, 0), (190, 0.7), (195, 0.5), (200, 1)\}$, where satisfaction counter-intuitively drops as throughput improves, clearly violating the fundamental monotonic property/meaning implied by the $P_1$ performance pattern.

\vspace{1em}

\noindent \textbf{Cases when \approach~does not work effectively:} There exists a small number of cases where \approach~is less effective, such as the requirement\footnote{Naturally, a higher FPS is better.}: \textit{``The video stream must maintain 60 FPS, but can drop to 30 FPS in power-saving mode, and should never exceed 120 FPS to save bandwidth.''} This interpretation of this requirement is logically a combination of at least $P_1$ and $P_2$ patterns. Since currently \approach~works on the given performance requirement with one pattern only each time, the above, although can be fragmented, might alter the original semantics such that \approach~cannot detect. In the above, the entire semantics implies that anything less than 60 FPS remains highly tolerable, but if we fragment only the first part, such a meaning becomes blurred, which causes \approach~to produce an initial draft that is far away from the stakeholder's true preferences (and hence might not be effectively quantified to be sufficiently close to the true preference even following the analogy reasoning and preference tuning).

In particular, a typical case we observed is related to the Retrieval-Analogical Preference Reasoning: when two performance requirements are semantically similar, but stakeholders’ preferences differ drastically. An example could be:

\begin{itemize}
    \item \textbf{Sample 1:}
    \begin{itemize}
        \item \textbf{Performance requirement:} ``\textit{In the Online Bookstore System, the search results for book titles shall be returned to the user within 5 seconds to ensure a smooth browsing experience.}''
\item \textbf{Initial quantification:} $f_{1,0}=(5.0,1.0),(5.5,0.0)$
\item \textbf{Ideal quantification:} $f^*_{1}=(5.0,1.0),(6.05,0.0)$
    \end{itemize}

        \item \textbf{Sample 2:}
    \begin{itemize}
        \item \textbf{Performance requirement:} ``\textit{In the Nuclear Power Plant Monitoring System, the status feedback for reactor cooling valves shall be returned to the console within 5 seconds to ensure real-time safety tracking.}''
\item \textbf{Initial quantification:} $f_{2,0}=(5.0,1.0),(5.5,0.0)$
\item \textbf{Ideal quantification:} $f^*_{2}=(5.0,1.0),(5.1,0.0)$
    \end{itemize}
\end{itemize}

Here, \textbf{sample 1} is chosen as the example, but using its transition makes the initial of \textbf{sample 2} even more distant from its ideal quantification.

\section{Full Results of the Sensitivity to Interaction Rounds $N$}
\label{sec:full_sen}

Table~\ref{tab:n_sensitivity_res} presents the detailed experimental results of the sensitivity analysis on the number of interactions $N$ with respect to each dataset.

\section{Full Results of the Sensitivity to $\Delta$ in Retrieval-Generative Quantification}
\label{sec:full_delta_sen}

Table~\ref{tab:delta_sensitivity} presents the detailed experimental results of the sensitivity analysis on the value of $\Delta$ with respect to each dataset.

\end{document}